\documentclass[apl,superscriptaddress,reprint,amssymb,aps,floatfix,showkeys]{revtex4-1}
\usepackage{amsmath}%
\usepackage{amsfonts}%
\usepackage{amssymb}%
\usepackage{graphicx}
\usepackage{color}
\usepackage{tabularx}
\usepackage{braket}
\usepackage{mathrsfs}
\usepackage{adjustbox}

\usepackage{array} 

\usepackage{makecell}

\usepackage[normalem]{ulem}  % For \sout{}
\usepackage{xcolor}          % For \textcolor{}

\begin{document}

\title{Inverse Design of Three-Dimensional Microwave Cavities for Optimizing Electromagnetic Helicity}

\author{Emma Paterson}
\affiliation{Quantum Technologies and Dark Matter Labs, Department of Physics, University of Western Australia, Crawley, Perth 6009, Australia}

\author{Jeremy Bourhill}
\affiliation{Quantum Technologies and Dark Matter Labs, Department of Physics, University of Western Australia, Crawley, Perth 6009, Australia}

\author{Maxim Goryachev}
\email{maxim.goryachev@uwa.edu.au}
\affiliation{Quantum Technologies and Dark Matter Labs, Department of Physics, University of Western Australia, Crawley, Perth 6009, Australia}

\begin{abstract}
We present an inverse-design framework for engineering three-dimensional microwave cavity resonators that support modes with nonzero electromagnetic helicity. Helicity maximisation is formulated as a boundary-shape optimisation problem, enabling systematic exploration of prescribed boundary-shape parameter spaces and the identification of high-helicity configurations for which the required combination of boundary parameters is difficult to anticipate a priori through heuristic design alone. We applied this framework to several cavity families composed of smooth, edge-free components, including globally twisted cavities with control-point-defined cross-sections realised in both linear and ring configurations, cavities defined by the intersection of orthogonal prisms, sphere-subtracted cylindrical cavities, and parametrised surface resonators. Two gradient-free optimisation strategies, a genetic algorithm and Bayesian optimisation, were independently employed to explore compact sets of design parameters for these geometries and to optimise a scaled-helicity figure of merit for the dominant helical mode, evaluated via finite-element eigenmode analysis. Robustness to manufacturing tolerances was quantified by applying Gaussian geometric perturbations to the optimised cavities and evaluating statistical robustness metrics that penalise sensitivity to geometric variation. The optimisation reveals clear physical design principles governing the generation of high electromagnetic helicity in three-dimensional microwave cavities.
\end{abstract}

\date{\today}
\maketitle

\renewcommand{\thesubsection}{\thesection.\arabic{subsection}}
\renewcommand{\thesection}{\arabic{section}}
\renewcommand{\thefigure}{\arabic{figure}}
\renewcommand{\thetable}{\arabic{table}}

\setcounter{secnumdepth}{3} % 1=section, 2=subsection, 3=subsubsection

\renewcommand{\thesubsubsection}{\thesection.\arabic{subsection}.\arabic{subsubsection}}

\makeatletter
\gdef\p@subsubsection{\thesubsection.} % so \ref prints 2.2.1 not just 1
\makeatother

\makeatletter
\gdef\p@subsection{}%
\gdef\p@subsubsection{}%
\makeatother

\section*{Introduction}
Inverse design methodologies, originally developed for nanophotonic devices, have transformed how electromagnetic structures are conceived by shifting from intuition-driven layouts to algorithmically discovered geometries that optimise prescribed performance metrics with minimal human intervention~\cite{Kim2025,Wang2021Intelligent,Molesky_2018,Wiecha2022,Schul2026}. This framework has enabled the realisation of a broad range of photonic devices, including wavelength demultiplexers~\cite{Inverse_Design_Demultiplexer_Su_2018,Piggott2017FabricationConstrained}, beam-steering metasurfaces~\cite{Beam_Stearing_ID_Thureja_2020}, and photonic-crystal circuits~\cite{InverseDesigned_2021_Vercruysse}. More recently, inverse design and topology optimisation have also been applied to microwave components, including filters, power splitters, and antennas~\cite{ZolfagharyPour2026,Morrison2025Rapid}.

In the present work, the optimisation objective is electromagnetic helicity,
$\mathscr{H}$, defined as the projection of electromagnetic spin angular
momentum onto its linear momentum~\cite{Alpeggiani_Electromagnetic_2018,
PhysRevLett.113.033601,Bliokh_2013}. Electromagnetic helicity therefore
provides a direct measure of the handedness, or chirality, of an
electromagnetic field. For a resonant mode $i$, we quantify its global
chirality using the dimensionless total helicity $\mathscr{H}_i$, with the
full normalisation defined in Sec.~\ref{sec:Fitness_Eval}. 

Heuristically designed microwave cavity resonators engineered to support large $|\mathscr{H}_i|$ have previously been reported~\cite{Bourhill_twisted_anyon_cavity_2023,paterson2025electromagnetic,paterson2025dynamicallytuneablehelicitytwisted,paterson_Berry}. These resonators consist of prisms with a $D_n$ cross-section whose spatial symmetry is broken by introducing a uniform twist along the prism height. The resulting mirror-asymmetric boundary conditions give rise to eigenmodes with nonzero $|\mathscr{H}_i|$ through two distinct mechanisms: hybridisation of near-degenerate transverse-electric (TE) and transverse-magnetic (TM) modes, and self-interference of TE modes. Irrespective of the mechanism by which a helical mode arises, $|\mathscr{H}_i|$ is determined by the degree of non-orthogonality between the $\mathbf{E}_i(\mathbf{r})$ and $\mathbf{H}_i(\mathbf{r})$ fields coexisting throughout the cavity volume.

While twisting provides one route to introducing mirror-asymmetric boundary conditions that support helical electromagnetic modes, alternative ways of introducing mirror asymmetry may yield cavity geometries that support greater $|\mathscr{H}_i|$ or offer improved performance in other relevant metrics, including robustness to geometric perturbations and cavity quality factor. This motivates the exploration of a broader range of mirror-asymmetric three-dimensional (3D) cavity constructions.

Unlike the previously studied twisted-prism resonators, whose design space can be explored heuristically by varying comparatively simple geometric features such as the twist angle and cross-sectional symmetry, more general mirror-asymmetric cavity boundaries do not necessarily admit an equally simple design principle for increasing $|\mathscr{H}_i|$. It is not generally apparent \textit{a priori} how the parameters defining these cavity boundaries should be varied, either individually or collectively, to increase helicity. Even small geometric variations can substantially reshape the $\mathbf{E}_i(\mathbf{r})$ and $\mathbf{H}_i(\mathbf{r})$ field distributions and alter both the frequencies and identities of the supported eigenmodes. Consequently, there is no simple mapping between individual boundary-shape parameters and $|\mathscr{H}_i|$, making it difficult to identify high-helicity configurations from physical intuition alone. 

This optimisation problem is therefore a natural candidate for inverse design, with $|\mathscr{H}_i|$ providing a figure of merit (FoM) for evaluating the supported eigenmodes as the geometric parameters defining the cavity boundary are varied, irrespective of their field structure or modal identity. However, the problem considered here differs from many established inverse-design formulations in several important respects.

Inverse-design problems are commonly solved using adjoint-based optimisation, in which forward and adjoint solves are used to compute gradients of the FoM with respect to the design variables~\cite{Hughes_2018,Deng2022}. These gradients enable efficient optimisation in high-dimensional parameter spaces. Such approaches can optimise electromagnetic responses to prescribed excitations using frequency- or time-domain simulations~\cite{Hughes_2018,Gedeon2023,Hammond2022}. Gradient-based optimisation has also been successfully applied to cavity geometries through eigenmode studies~\cite{Ziegler2024GradientBased,Flisgen2006Eigenvalue}. In such formulations, mode-matching techniques can be used to track a prescribed eigenmode as the geometry evolves~\cite{Ziegler2024GradientBased}.

The present work instead uses eigenmode studies to evaluate the intrinsic properties of the resonances supported by each candidate cavity geometry, while allowing the identity of the mode maximising the FoM to change throughout the optimisation. This formulation removes dependence on port geometry and excitation-dependent coupling, avoiding preferential selection of resonances that happen to couple strongly to a particular excitation. It also enables multiple eigenmodes within a broad microwave frequency window to be evaluated for each candidate geometry. The eigenmode formulation additionally limits computational cost by avoiding driven frequency sweeps that would otherwise be required to resolve multiple resonances across the specified frequency window for every candidate geometry.

For each candidate geometry, the FoM is evaluated for each of the calculated eigenmodes within this frequency window, and the mode yielding the largest $|\mathscr{H}_i|$ is selected. Unlike eigenmode optimisation in which a prescribed mode is continuously tracked, no single modal identity is imposed here. As the cavity geometry evolves, the ordering and field structure of the supported eigenmodes may change, allowing the mode maximising the FoM to switch between physically distinct resonances.

This distinction also motivates the choice of optimisation strategy. Although the properties of an individual modal branch may vary smoothly with the design parameters, changes in the identity of the maximising mode can introduce transitions at which the overall optimisation objective may not be smoothly differentiable. Gradient-free methods are therefore well suited to the present formulation because they allow candidate geometries to be compared directly while permitting the modal identity associated with the maximum FoM to change as the geometry evolves. We employ two such strategies: a genetic algorithm (GA) and Bayesian optimisation (BO)~\cite{Molesky_2018,Wang2021Intelligent,Kim2025}.

While inverse-design methods have been widely applied to driven electromagnetic problems and to the optimisation of prescribed cavity eigenmodes, comparatively little attention has been given to fully three-dimensional metallic microwave cavity boundaries without exploitable geometric symmetry, particularly when the optimisation is not constrained to a single prescribed and continuously tracked eigenmode. The present work addresses this regime using distinct cavity families, representing physically motivated constructions chosen to introduce mirror asymmetry and defined by compact parameterisations of their three-dimensional boundaries. Gradient-free optimisation methods are then used to explore the prescribed boundary-shape parameter space within each cavity family and identify configurations that maximise $|\mathscr{H}_i|$. This work establishes a practical inverse-design framework for systematically optimising prescribed three-dimensional microwave cavity boundary-shape parameter spaces with respect to electromagnetic helicity.

The ability to systematically engineer microwave cavity modes with large $|\mathscr{H}_i|$ has applications in systems involving interactions with chiral matter and pseudoscalar sources, including enantioselective spectroscopy~\cite{Tang2011,Hendry:2010ug}, enhanced circular dichroism for chiral sensing and material characterisation~\cite{Multunas_Circular_Dichroism_Crystals}, and resonant axion dark matter searches~\cite{Invisible_Axion_Search_2021_Sikivie,paterson2025electromagnetic}. Possible future applications include chiral quantum sensing, where helicity-engineered modes may enable spin-selective interactions with defect centres such as nitrogen-vacancy (NV) centres~\cite{Both2022,Toward_quantum_sensing_Volker_2023}, and nonreciprocal microwave devices exploiting time-reversal-symmetry breaking~\cite{UpAndComing_Advances_Kutsaev_Advanced_Photonics_2021}.

This paper is organised as follows. Sec.~\ref{sec:Inverse-Challenges-Opportunities} outlines the principal challenges associated with the inverse design of fully 3D microwave cavities. Sec.~\ref{sec:methods} describes the optimisation and robustness-assessment framework, and Sec.~\ref{sec:Geometries} applies this framework to several parameterised cavity families. 

\section{Inverse Design of Microwave Cavities: Challenges and Opportunities}~\label{sec:Inverse-Challenges-Opportunities}

\subsection{Challenges of Inverse Design in Fully 3D Cavities}

While inverse design and topology optimisation are well established for planar photonic systems~\cite{Molesky_2018,Wiecha2022,Wang2021Intelligent}, their application to fully 3D microwave cavities introduces additional geometric and computational considerations. In photonics, many inverse-design problems exploit planar or quasi-two-dimensional (quasi-2D) layouts with symmetry axes, reducing the problem to 2D permittivity maps~\cite{Molesky_2018,Wiecha2022}. Dimensional-reduction techniques have also been applied to helically coiled optical waveguides, where the constant curvature and torsion of the helical geometry permit reformulation of the 3D Helmholtz eigenproblem as an equivalent 2D eigenproblem on the waveguide cross-section~\cite{Gopalakrishnan2024}. By contrast, inverse design on truly 3D microwave cavity resonators with no exploitable symmetries requires parametrising fully 3D metal boundaries, a more complex task than optimising a 2D geometry.

A further challenge is the substantially greater computational cost associated with volumetric finite-element simulations of fully 3D field distributions. Compared with planar or quasi-2D photonic inverse-design problems, these simulations require substantially larger computational meshes and correspondingly greater memory and computational resources~\cite{Kang2024LargeScale}. Moreover, accurately resolving the spatial distribution of $h_i(\vec{r})$ further increases the computational burden.

Despite these challenges, inverse design has been successfully applied to enhance the chirality of fully 3D nanostructures~\cite{Taguchi2025}. Applying these techniques to metallic microwave cavity resonators, however, introduces different design considerations: one must account for realistic metal losses, enforce minimum feature sizes along all axes, avoid unsupported thin walls or overhangs, and accurately capture curved conducting boundaries~\cite{Khoram2020,ZolfagharyPour2026}. 

\subsection{Fabrication Constraints due to Additive Manufacturing}

Beyond computational and modelling challenges, experimental realisation of geometrically complex 3D microwave cavities presents additional difficulties. Such structures are most readily fabricated using metal additive manufacturing (AM), which provides substantial geometric freedom but can introduce dimensional deviations and surface defects arising from layer-wise fabrication, thermal distortion, partially melted powder, residual stresses, and the stair-step effect on curved surfaces~\cite{chaghazardi2022review,morphological_Hamidi_2018,Microstructures_Rafi_2013,Melia_2020}. These effects can increase surface roughness and produce local geometric deviations from the intended cavity boundary, both of which are detrimental to microwave cavity performance~\cite{surface_roughness_Giovanni,Wi_2019_Electropolishing,Tyagi2018}.

Post-processing methods such as electropolishing can reduce surface roughness, but their effectiveness is limited in geometries containing sharp edges, deep recesses, and strongly varying curvature, where non-uniform current distribution and restricted electrolyte access can produce uneven material removal~\cite{LANDOLT19871,Aryan2015,PEREZ2015352,TYAGI201932}. These fabrication and post-processing constraints motivate the use of smooth, edge-free (EF) cavity boundaries in the present inverse-design framework and the explicit assessment of each optimised geometry's sensitivity to dimensional perturbations.

\subsection{Optimisation Design and Robustness Assessment}

\begin{figure*}[t]
    \centering
    \includegraphics[width=\textwidth]{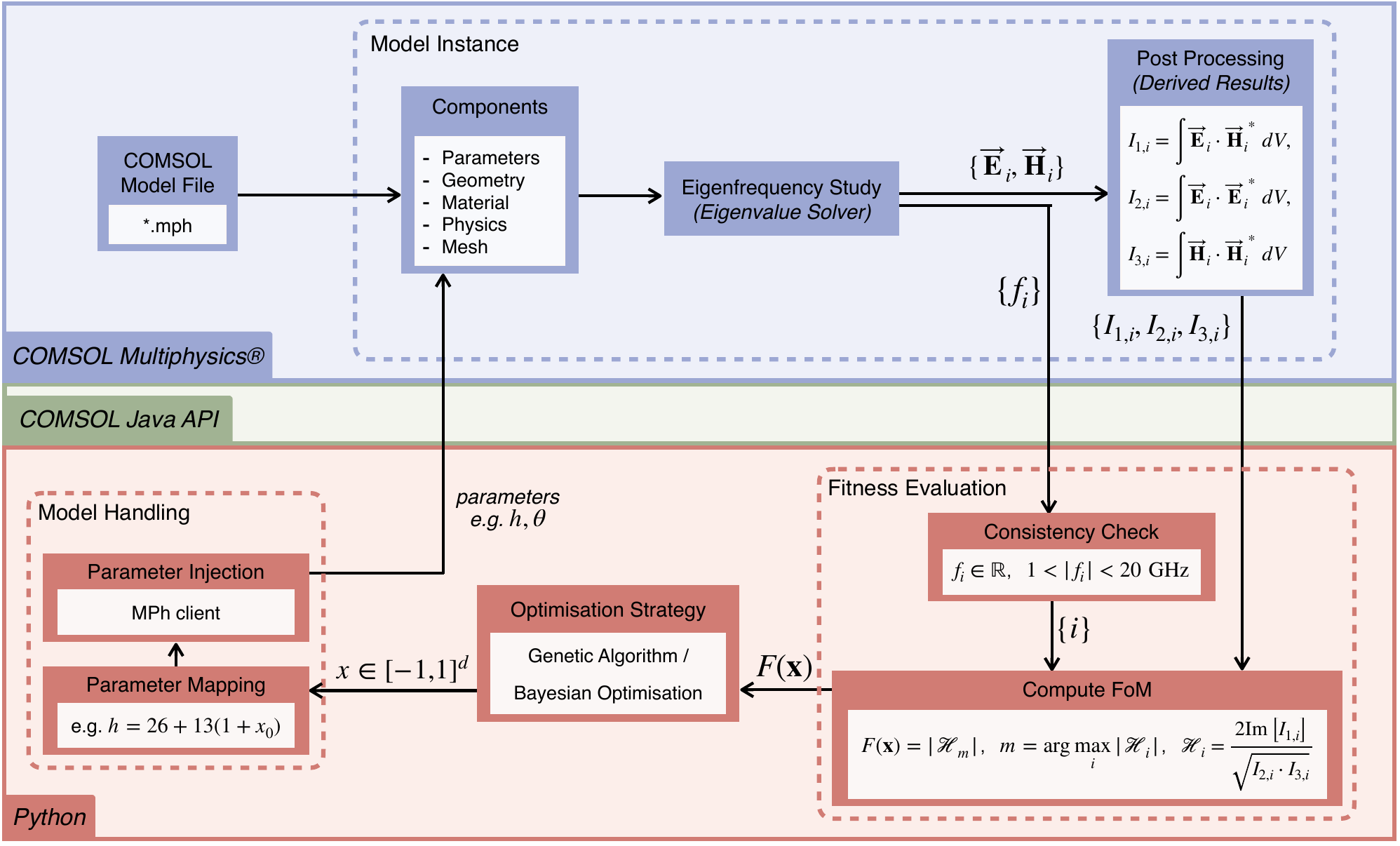}
    \caption{Block diagram of the inverse design framework used in this work to optimise microwave cavity geometries for electromagnetic helicity, $\mathscr{H}$. A COMSOL Multiphysics\textsuperscript{\textregistered}~\cite{comsol_multiphysics_v6.4} model file containing the parameterised cavity geometry, material properties, physics interfaces, and mesh is loaded into a COMSOL model instance via an MPh~\cite{MPH} client using the COMSOL Java API. A normalised design vector $\mathbf{x}\in[-1,1]^d$, generated by the optimisation algorithm, is mapped to physical geometric parameters and injected into the model. An eigenfrequency study is executed, yielding a set of cavity eigenmodes $\{i\}$, where $i$ indexes individual eigenmodes, with corresponding eigenfrequencies $\{f_i\}$ and $k$ number of post-processing volume integrals $\{I_{k,i}\}$. For eigenmodes with real $f_i \in [1,20]~\mathrm{GHz}$, the FoM $F(\mathbf{x}) = |\mathscr{H}_m|$, where $m$ denotes the most helical mode in the retained set, is computed and returned to the optimisation routine, closing the optimisation loop. }
    \label{fig:block_diagram}
\end{figure*}

The microwave $Q$-factor is highly sensitive to surface smoothness and conductivity~\cite{Creedon2016}, making uniformly smooth internal surfaces in additively manufactured cavities a central challenge for realizing low-loss, high-$Q$ resonators. To address these limitations, our inverse-design optimisation is carried out over 3D cavity families constructed from smooth, EF components. By ensuring that the constituent components avoid sharp features, corners, and abrupt curvature variations, the resulting cavities are compatible with additive manufacturing and suitable for electropolishing during post-processing. 

We make this optimisation tractable by restricting the search to compact geometric parameterisations. For each candidate geometry, we evaluate only a small, fixed set of eigenmodes. We also tune the PyGAD optimisation parameters to minimise the total number of finite-element eigenmode solves required.

Manufacturability alone is insufficient; practical cavity designs must also remain robust to the geometric variations introduced during fabrication. To this end, we evaluate the performance of each optimised design under random Gaussian geometric perturbations, identifying cavities that retain high $|\mathscr{H}_i|$ under parameter variations and are therefore less sensitive to dimensional deviations arising from electropolishing and the intrinsic fabrication tolerances of metal additive-manufactured components. Our robustness-evaluation framework thereby reduces an otherwise intractable exploration of a high-dimensional tolerance space to a small set of interpretable statistical metrics.

\section{Methods}~\label{sec:methods}

Figure~\ref{fig:block_diagram} presents a schematic overview of the inverse design framework. The diagram illustrates how Python-based optimisation routines interface with COMSOL Multiphysics\textsuperscript{\textregistered}~\cite{comsol_multiphysics_v6.4} to perform electromagnetic finite-element simulations of the cavity resonators. The components of this framework are described in detail below.

\subsection{Fitness Evaluation}
\label{sec:Fitness_Eval}

For each cavity family investigated, a dedicated COMSOL Multiphysics\textsuperscript{\textregistered} model is constructed. Each model contains a fully parameterised cavity geometry defined by a set of global geometric parameters, with perfect electric conductor boundary conditions imposed on all metallic surfaces. The cavity volume is assigned free-space material properties, with relative permittivity $\epsilon_r=1$, relative permeability $\mu_r=1$, and electrical conductivity $\sigma=0~\mathrm{S\,m^{-1}}$. The Electromagnetic Waves, Frequency Domain physics interface is employed to model the electromagnetic fields. A free tetrahedral mesh is used to resolve the full 3D electromagnetic field structure. Each model additionally contains a predefined eigenfrequency study and the post-processing volume integrals required to evaluate the electromagnetic helicity of the supported eigenmodes.

For each resonant mode $i$, the local time-averaged helicity density is evaluated as~\cite{paterson2025electromagnetic}
\begin{equation}
    h_i(\mathbf{r})
    =
    \frac{2}{V}\,
    \mathrm{Im}\!\left[
        \mathbf{e}_i(\mathbf{r})
        \cdot
        \mathbf{h}_i^{*}(\mathbf{r})
    \right]
    =
    \frac{
        2\,\mathrm{Im}\!\left[
            \mathbf{E}_i(\mathbf{r})
            \cdot
            \mathbf{H}_i^{*}(\mathbf{r})
        \right]
    }{
        V\sqrt{\mathcal{E}_i\mathcal{M}_i}
    },
    \label{eq:local_helicity}
\end{equation}
where $\mathbf{E}_i(\mathbf{r})$ and $\mathbf{H}_i(\mathbf{r})$ are the electric and magnetic fields of the resonant mode and $V$ is the cavity volume. The quantities
\begin{equation}
    \mathcal{E}_i
    =
    \frac{1}{V}
    \int
    \mathbf{E}_i^{*}(\mathbf{r})
    \cdot
    \mathbf{E}_i(\mathbf{r})\,dV,
    \qquad
    \mathcal{M}_i
    =
    \frac{1}{V}
    \int
    \mathbf{H}_i^{*}(\mathbf{r})
    \cdot
    \mathbf{H}_i(\mathbf{r})\,dV
    \label{eq:field_normalisation}
\end{equation}
are the volume-averaged squared electric and magnetic field amplitudes, respectively. The corresponding root-mean-square-normalised fields are
\begin{equation}
    \mathbf{e}_i(\mathbf{r})
    =
    \frac{
        \mathbf{E}_i(\mathbf{r})
    }{
        \sqrt{\mathcal{E}_i}
    },
    \qquad
    \mathbf{h}_i(\mathbf{r})
    =
    \frac{
        \mathbf{H}_i(\mathbf{r})
    }{
        \sqrt{\mathcal{M}_i}
    }.
    \label{eq:rms_fields}
\end{equation}
Physically, a large magnitude of $h_i(\mathbf{r})$ corresponds to a strong local overlap between parallel components of the electric and magnetic fields and hence to a well-defined local electromagnetic handedness.

The total electromagnetic helicity of the resonant mode $i$ is obtained by integrating the local helicity density over the cavity volume,
\begin{equation}
    \mathscr{H}_i
    =
    \int
    h_i(\mathbf{r})\,dV,
    \label{eq:total_helicity}
\end{equation}
which provides a quantitative measure of the global chirality of the electromagnetic mode. Under the normalisation defined above, $\mathscr{H}_i$ is dimensionless. By the Cauchy--Schwarz inequality, $|\mathscr{H}_i/2| \leq 1$, implying that $|\mathscr{H}_i| \leq 2$, with the limiting values $\mathscr{H}_i=\pm2$ corresponding to maximally helical fields. Throughout this work, the optimisation targets the magnitude $|\mathscr{H}_i|$ rather than its sign. If an application requires the opposite handedness, the cavity can be realised as the mirror image of the optimised geometry.

Programmatic control of the COMSOL models is achieved using the Python package MPh~\cite{MPH}, which provides an interface to the COMSOL Model Java API. Through MPh, a client is created that launches a COMSOL session and loads the predefined model file. For each candidate geometry, the optimisation routine generates a normalised design vector $\mathbf{x} \in [-1,1]^d$, where $d$ is the number of geometric parameters being optimised. Each component of $\mathbf{x}$ is affinely mapped to the corresponding physical parameter in the COMSOL model, such as cavity height, twist angle, or radial control points, according to the parameterisation of the cavity family under consideration. The admissible ranges of these physical parameters are selected to satisfy the geometric and fabrication constraints defined for each cavity family. Expressing the design variables in terms of $\mathbf{x}$ yields a compact, dimensionless search space that is well suited to optimisation. 

The mapped parameters are injected into the COMSOL model via the Java API, and an eigenfrequency study is executed to compute the cavity eigenmodes. The number of eigenmodes solved for is chosen on a per-geometry basis to ensure adequate sampling of the spectrum while maintaining reasonable solve times. Using this API, the corresponding eigenfrequencies and post-processed volume integrals required to evaluate the electromagnetic helicity are extracted. Only modes with real eigenfrequencies in the range $f_i\in[1,20]~\mathrm{GHz}$ are retained; all others are discarded.

The FoM for each candidate geometry is then computed as
\begin{equation}
    F(\mathbf{x}) = |\mathscr{H}_m|,
    \qquad
    m = \operatorname*{arg\,max}_{i} |\mathscr{H}_i|,
\end{equation}
where the maximisation is performed over the retained eigenmodes and $m$ denotes the index of the mode with the largest magnitude of electromagnetic helicity. Importantly, $m$ is determined independently for each candidate geometry; no single eigenmode is prescribed or continuously tracked throughout the optimisation. The mode contributing to the fitness may therefore change in modal identity and field structure as the cavity geometry evolves. The resulting value $F(\mathbf{x})$ is then supplied to the optimisation algorithm as the fitness value.

\begin{table*}[t]
\centering
\caption{Genetic-algorithm parameters used for the optimisation of the cavity geometries.}
\label{tab:pygad_parameters}
\setlength{\tabcolsep}{5pt}
\renewcommand{\arraystretch}{1.1}
\begin{tabular}{lcccccccc}
Geometry &
$N_{\!gen}$ &
$N_{\!par}$ &
$N_{\!pop}$ &
$N_{\!elite}$ &
$d$ &
$f_{\!mut}$ &
$\sigma_{\!mut}$ &
$\mathcal{T}_{\mathrm{stop}}$ \\
\hline
EF twisted linear (baseline) [Sec.~\ref{sec:EF-Tw-WG}] &
99 & 9 & 44 & 10 & 10 & 0.10 & 0.18 &
saturate\_98, saturate\_7 \\
EF twisted linear (expanded $r_i$) &
200 & 56 & 140 & 6 & 10 & 0.35 & 0.18 &
saturate\_40 \\
EF twisted ring [Sec.~\ref{sec:EF-Tw-RING-WG}] &
260 & 56 & 140 & 6 & 10 & 0.35 & 0.08 &
saturate\_40 \\
Orthogonal prism-intersection [Sec.~\ref{sec:Cross_Cyl}] &
123 & 9 & 47 & 10 & 16 & 0.10 & 0.18 &
reach\_98, saturate\_19 \\
Sphere-subtracted cylindrical [Sec.~\ref{sec:sculpture_cavity}] &
199 & 11 & 49 & 10 & 26 & 0.10 & 0.08 &
reach\_98, saturate\_7 \\
Parametrised surface [Sec.~\ref{sec:param_surface}] &
100 & 11 & 30 & 6 & 6 & 0.17 & 0.08 &
reach\_98, saturate\_7 \\
\end{tabular}
\end{table*}

\subsection{Optimisation Strategies}~\label{sec:optimisation_strategies}

\subsubsection{Genetic Algorithm}~\label{sec:gen_alg}

A genetic algorithm (GA) is one of the optimisation strategies employed to maximise the FoM. This population-based evolutionary algorithm iteratively refines the geometry parameters through a process of selection, crossover, and mutation. At each generation, a new population of candidate geometries is generated, evaluated, and ranked according to $F(\mathbf{x})$. This process continues until convergence is reached, yielding a cavity shape that maximises the $\mathscr{H}$-based FoM. 

The GA is implemented using the PyGAD library~\cite{gad2023pygad} with tournament selection, elitism, and Gaussian mutation. A single-point crossover is applied to combine parent solutions, while Gaussian mutation with a standard deviation of $\sigma_{\mathrm{mut}} = 0.08\,\text{--}\,0.18$ (depending on the cavity family) perturbs a geometry-dependent fraction $f_{\rm mut}=0.10-0.35$ of the genes in each generation. Mutations are clipped to $[-1,1]$ in the normalised parameter space to ensure that all design variables remain within their physical bounds. Non-convergent COMSOL simulations are penalised with a large negative fitness value to prevent invalid solutions from propagating through the population.

Key GA parameters for all optimised cavities are listed in Table~\ref{tab:pygad_parameters}. The population size ($N_{\mathrm{pop}}$), number of generations ($N_{\mathrm{gen}}$), number of parents ($N_{\mathrm{par}}$), elitism count ($N_{\mathrm{elite}}$), and mutation rate ($f_{\mathrm{mut}}$) were scaled with $d$, to maintain adequate search coverage and genetic diversity. Convergence was monitored using the stopping criterion, $\mathcal{T}_{\mathrm{stop}}$, which terminated the optimisation once the fitness improvement plateaued. Detailed heuristic ranges for these parameters are provided in Appendix~\ref{sec:GAHeuristics}.

\subsubsection{Bayesian Optimisation}~\label{sec:bay_opt}

An alternative global optimisation strategy, Bayesian optimisation (BO), is used to search the design space by using results from COMSOL simulations to train a Gaussian Process (GP) surrogate model of $F(\mathbf{x})$ across the design space. This approach can reduce the number of required simulations by using the surrogate model to guide evaluations towards promising regions of the parameter space.

The GP uses a Matérn covariance kernel with smoothness parameter $\nu = 2.5$, implemented via the \texttt{GaussianProcessRegressor} in \texttt{scikit-learn}~\cite{scikit-learn}, with automatic optimisation of the kernel hyperparameters (signal variance and length scale). The Matérn kernel defines how the correlation between two design points decays with distance in parameter space and determines the smoothness of the surrogate function. Automatic hyperparameter optimisation ensures that the kernel parameters adapt to the scale and variability of the underlying COMSOL data at each iteration. 

An initial set of $N_{\mathrm{init}}=3000$ points is sampled uniformly within the normalised design space ($\mathbf{x}$) and evaluated in COMSOL to form a training dataset $\{\mathbf{x}_i, F(\mathbf{x}_i)\}$. The GP is then trained to predict both the mean and uncertainty of $F(\mathbf{x})$ at untested geometries. New candidates are selected by maximising the Expected Improvement (EI) acquisition function, which balances exploration of uncertain regions and exploitation of high-performing areas. EI maximisation is performed using the L-BFGS-B algorithm with multiple random restarts, $N_{restart}=10$, within the same bounds.

At each iteration, one new geometry is selected, evaluated using COMSOL, and added to the dataset. The GP is retrained with the updated data, and the process repeats until the maximum number of iterations ($N_{\mathrm{BO}} = 200$) is reached. The best-performing BO design corresponds to the highest $F(\mathbf{x})$ observed over all evaluations. The GP predictive standard deviation, $\sigma_{\text{pred}}$, is analysed over $N_{\mathrm{test}}=2000$ random samples in $\mathbf{x}$ to quantify model confidence and coverage of the explored design space.

While both GA and BO optimise over the same geometric parameter set and use the same $F(\mathbf{x})$, they differ fundamentally in their search strategies: GA explores the space through stochastic sampling and evolutionary pressure, whereas BO constructs a statistical model of the fitness landscape to target regions of high performance. Both GA and BO are employed here as complementary search strategies for exploring high-dimensional geometric parameter spaces. The choice of optimisation method for each cavity family reflects practical considerations such as computational cost and parameter-space complexity. 

\subsection{Mesh Convergence}~\label{sec:mesh_convergence}

Because the cavity geometry evolves throughout the optimisation, a new free tetrahedral mesh is generated by COMSOL for each candidate geometry using a prescribed set of mesh-generation parameters for the corresponding cavity family. These parameters are held fixed throughout the optimisation and are reported in the respective cavity-family subsections of Sec.~\ref{sec:Geometries}. Following completion of the optimisation, mesh-convergence testing is performed on the final optimised cavity from each family to verify that the reported value of $|\mathscr{H}_m|$ is not an artefact of mesh discretisation.

Mesh convergence was assessed via a parametric sweep of the maximum mesh element size, confirming that the FoM stabilised under successive refinement. Convergence is quantified by evaluating the relative change in $F(\mathbf{x})$ between consecutive meshes and verifying that it falls below a $0.001\%$ threshold. This procedure verifies that the reported FoM of each final optimised cavity is insensitive to mesh size.

\subsection{Optimised Cavity Performance Metrics}~\label{sec:opt_cavity_perf_metrics}

The performance of the optimised cavity from each cavity family is characterised using a set of key metrics, including the resonant frequency $f_i$, $|\mathscr{H}_i|$, surface-loss sensitivity, and robustness to perturbations. To enable fair comparison between distinct cavity families, all models are uniformly scaled to maintain a constant surface-to-volume ratio prior to eigenmode analysis.

Sensitivity to surface losses is quantified by the geometric factor, $G$, which characterises the sensitivity of a resonant mode to dissipation at the conducting walls and is defined for a given mode $i$ as~\cite{Krupka2005}:
\begin{equation}
        G_i = 2\pi f_i \frac{\displaystyle \iiint_{V} \mu_0 |\vec{\mathbf{H}}_i|^2 \, dv}{\displaystyle \iint_{S} |\vec{\mathbf{H}}_i^\tau|^2 \, ds},
        \label{eq:geom_factor}
\end{equation}
where $\mu_0$ is the vacuum permeability. The volume integral is taken over the cavity volume $V$, while the surface integral evaluates the tangential magnetic field $\vec{\mathbf{H}}_i^\tau$ over the conducting walls. Physically, a larger G indicates reduced sensitivity to conductor surface losses, corresponding to a larger ratio of stored magnetic energy to the surface-integrated tangential magnetic-field intensity at the cavity walls, thereby reducing surface-induced dissipation for a given stored energy and surface resistance. $G$ is related to the intrinsic quality factor, $Q_0$, through~\cite{Creedon2016}:
\begin{equation}
        Q_0 = \frac{G}{R_S},
\end{equation}
where $R_S$ denotes the surface resistance. To enable comparison between cavity families, for which the most helical mode occurs at different resonant frequencies, we define the frequency-normalised geometric factor
\begin{equation}
\tilde{G}_i = \frac{G_i}{2\pi f_i \mu_0},
\label{eq:freq-norm-G-Factor}
\end{equation}
which isolates the purely geometric contribution to surface-loss sensitivity. 

\begin{table*}[t]
\centering
\caption{Parameters for the optimised cavities in the EF twisted family.}
\label{tab:optimised_linear_cavity}
\setlength{\tabcolsep}{5pt}
\renewcommand{\arraystretch}{1.1}
\begin{tabular}{lcccc}
Geometry & Optimisation & $H$ (mm) & $\theta$ (rad) &
$\{r_i\}$ (mm) \\
\hline
EF twisted linear (baseline) [Sec.~\ref{sec:EF-Tw-WG}] & GA & 48.6 & 1.94 & \{22.9, 17.9, 22.7, 31.6, 25.1, 19.2, 24.7, 30.9\} \\
 & BO & 44.5 & 1.96 & \{25.9, 29.0, 19.7, 18.6, 25.4, 28.3, 19.6, 19.6\} \\
EF twisted linear (expanded $r_i$) & GA & 35.2 & 0.716 & \{21.0, 40.5, 23.1, 20.4, 38.1, 19.9, 23.1, 39.8\} \\
EF twisted ring [Sec.~\ref{sec:EF-Tw-RING-WG}] & GA & 995 & $10\pi$ &
\{30.8, 17.4, 25.9, 19.5, 15.6, 29.0, 17.4, 16.5\} \\
\end{tabular}
\end{table*}

To evaluate how small perturbations in the high-dimensional parameter set defining the optimised cavity geometry affect $F(\mathbf{x})$, each parameter of the optimised geometry $\mu_j$ is treated as subject to independent Gaussian uncertainty with a prescribed standard deviation $\sigma_j$, representing a perturbation on the order of 1\% in the normalised design space. In practice, these uncertainties represent the fabrication tolerances inherent to AM parts and electropolishing, as discussed in the introduction. They also capture geometric errors that may arise from post-fabrication assembly effects, such as misalignment, assembly-induced strain or deformation, and probe-induced perturbations.

\begin{figure}[t]
     \begin{center}
            \includegraphics[width=0.45\textwidth]{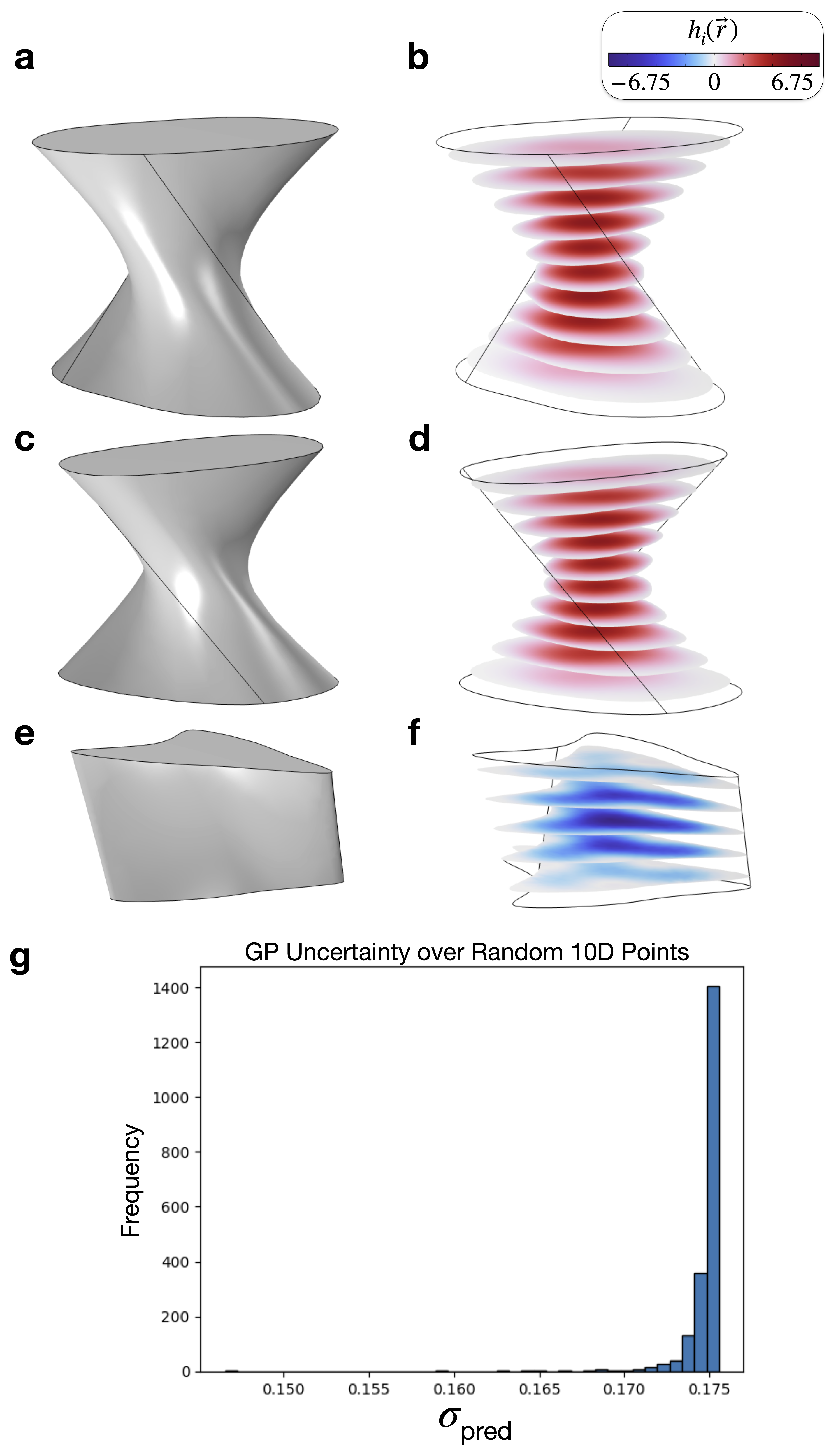}
            \end{center}
            \caption{(A) GA-optimised EF twisted linear cavity geometry and (B) its corresponding $h_i(\vec{r})$ field distribution. (C) BO-optimised cavity and (D) its corresponding $h_i(\vec{r})$ field distribution. (E) GA-optimised expanded-radius variant and (F) its corresponding $h_i(\vec{r})$ field distribution. (G) Histogram of $\sigma_{\text{pred}}$ across 2,000 random points in the 10D input space.}
    \label{fig:TWIST-T-Plots}
\end{figure}

\begin{table*}[t]
\centering
\caption{Statistical robustness metrics for the GA-optimised cavities under Gaussian perturbations.}
\label{tab:robustness_metrics}
\setlength{\tabcolsep}{5pt}
\renewcommand{\arraystretch}{1.1}
\begin{tabular}{l c c c c c}
Geometry & $\bar{F}$ & $\sigma_F$ & $F_{0.05}$ & $F_{\min}$ & $\mathcal{R}$ \\
\hline
EF twisted linear (baseline) [Sec.~\ref{sec:EF-Tw-WG}] & 1.02  & 0.0024 & 1.02  & 1.01  & 1.02  \\
EF twisted linear (expanded $r_i$) & 0.931 & 0.148  & 0.708 & 0.685 & 0.635 \\
EF twisted ring [Sec.~\ref{sec:EF-Tw-RING-WG}] & 1.44 & 0.0767 & 1.33 & 0.978 & 1.29 \\
Orthogonal prism-intersection [Sec.~\ref{sec:Cross_Cyl}] & 0.859 & 0.0620 & 0.691 & 0.621 & 0.735 \\
Sphere-subtracted cylindrical [Sec.~\ref{sec:sculpture_cavity}] & 0.867 & 0.0800 & 0.699 & 0.598 & 0.707 \\
Parametrised surface [Sec.~\ref{sec:param_surface}] & 0.495 & 0.191 & 0.283 & 0.279 & 0.113 \\
\end{tabular}
\end{table*}

Direct random sampling of these uncertainties would leave large gaps in the multidimensional tolerance space when only $N \approx 10^2$ model evaluations are feasible, so Latin Hypercube Sampling (LHS) was used to ensure even sampling across the uncertainty range of each nominal parameter $\mu_j$. Each uniform draw $u_{sj}\in(0,1)$ is mapped through the inverse normal cumulative distribution, $\Phi^{-1}(u_{sj})$, to generate perturbed parameter vectors $x_{sj} = \mu_j + \sigma_j \,\Phi^{-1}(u_{sj})$, producing an ensemble $\{\mathbf{x}_s\}_{s=1}^{N}$ drawn from $\mathcal{N}(\boldsymbol{\mu},\mathrm{diag}\,\boldsymbol{\sigma}^2)$ while sampling the design space. A fixed random seed (\texttt{seed=42}) was used to initialise the pseudorandom number generator that controls the LHS, allowing the same ensemble of perturbed geometries to be regenerated for reproducibility.

The FoM was evaluated for each perturbed geometry $\mathbf{x}_s$ in COMSOL, yielding a distribution of values $\{F(\mathbf{x}_s)\}$ that quantifies how machining tolerances influence performance. The mean $\bar{F}$ and standard deviation $\sigma_F$ describe the central tendency and variability, while the 5th percentile $F_{0.05}$ and the minimum value $F_{\min}$ indicate the near- and worst-case outcomes. A robustness index, $\mathcal{R} = \bar{F} - 2\sigma_F$, combines average performance and variability, penalising designs whose $F(\mathbf{x})$ fluctuates strongly under perturbation and providing a single measure of geometric tolerance. This framework enables direct comparison between competing geometries and a clear assessment of their resilience to fabrication imperfections.

\section{Geometries}~\label{sec:Geometries}

\subsection{Edge-Free Twisted Linear Cavity}~\label{sec:EF-Tw-WG}

The previously investigated twisted prism resonators~\cite{Bourhill_twisted_anyon_cavity_2023,paterson2025electromagnetic,paterson2025dynamicallytuneablehelicitytwisted,paterson_Berry}, which employ cross-sections with $D_n$ symmetry as discussed in the Introduction, were fabricated via additive manufacturing of twisted prism waveguides and subsequently terminated with separately manufactured electrically conducting endcaps to form cavities. However, the polygonal cross-sections of these designs introduce sharp edges and internal corners, and therefore inherit the surface-treatment limitations discussed in the Introduction.

To overcome these limitations, the present approach employs smoothly varying, closed 2D interpolation curves in lieu of polygonal profiles for the additively manufactured waveguide. Each curve is defined by $n=8$ control points whose Cartesian coordinates are $\bigl(r_i\cos\varphi_i,\;r_i\sin\varphi_i\bigr)$, with fixed angular positions $\varphi_i=2\pi i/n$ and radial distances $r_i$ treated as optimisation variables. The resulting planar shape is then lofted into three dimensions by extrusion over a height $H$, with a uniformly distributed total twist angle $\theta$ applied along the height. The free parameters defining the design space, along with their respective ranges, are therefore $\theta \in [0,2\pi]$, $H \in [26,156]~\mathrm{mm}$, and $r_i \in [15,33]~\mathrm{mm}$. 

The tetrahedral mesh used for this configuration employed a maximum element size of $2.53~\mathrm{mm}$ and a minimum element size of $0.108~\mathrm{mm}$. A maximum element growth rate of $1.35$ was used, together with a curvature factor of $0.3$. The narrow-region resolution was set to $0.85$.

The optimisation was performed using both the GA and BO frameworks. For each candidate geometry, an eigenvalue analysis was carried out to compute $44$ eigenfrequencies in the vicinity of $10$~GHz, from which the values of $|\mathscr{H}_i|$ for the supported modes were evaluated.

The parameters of the optimised geometries are listed in Table~\ref{tab:optimised_linear_cavity}. The GA-optimised geometry and the corresponding $h_i(\vec{r})$ field distribution of its resonant mode are shown in Fig.~\ref{fig:TWIST-T-Plots}(a) and (b), respectively, yielding $\tilde{G}_i = 5.26~\mathrm{mm}$ and $\mathscr{H}_i = 1.03$. The BO-optimised counterpart, shown in Fig.~\ref{fig:TWIST-T-Plots}(c) and (d), achieves $\tilde{G}_i = 4.57~\mathrm{mm}$ and $\mathscr{H}_i = 1.02$. Both optimisation frameworks yield cavities that converge to similarly high $|\mathscr{H}_i|$ values, approaching the upper bound attainable within this design space.

Expanding the admissible range of the control-point radii to $r_i \in [5,59]~\mathrm{mm}$ yields a second GA-optimised EF twisted linear cavity that achieves $\mathscr{H}_i = 1.100$ and $\tilde{G}_i = 6.32~\mathrm{mm}$ (see Table~\ref{tab:optimised_linear_cavity} for the optimised cavity parameters). The optimised geometry and its corresponding $h_i(\vec{r})$ distribution are shown in Fig.~\ref{fig:TWIST-T-Plots}(e) and (f), respectively.

The statistical robustness metrics for the GA-optimised EF twisted linear cavity, optimised over the baseline and expanded-$r_i$ design spaces, are summarised in the first two rows of Table~\ref{tab:robustness_metrics}. The baseline geometry exhibits a very low $\sigma_F$ and a near-unity $\mathcal{R}$, demonstrating excellent stability of $|\mathscr{H}_i|$ under parametric perturbations. By contrast, the expanded-radius geometry attains higher $|\mathscr{H}_i|$ and $\tilde{G}_i$ at the expense of a substantially broader fitness distribution and reduced $\mathcal{R}$, reflecting increased sensitivity to geometric perturbations arising from large radial excursions and revealing a clear trade-off between maximising the degree of $|\mathscr{H}_i|$ supported and maintaining robustness to fabrication or alignment variations. 

For the BO-optimised surrogate model, predictive uncertainty was evaluated using the trained GP. The histogram in Fig.~\ref{fig:TWIST-T-Plots}(g) shows the distribution of $\sigma_{\text{pred}}$, across 2,000 random points in the 10D design space, which is sharply peaked around 0.175. This indicates that the GP is confident and well-calibrated throughout the explored region, reinforcing the reliability of the BO-generated solution. 

Both the GA and BO approaches converge to near-identical $|\mathscr{H}_i|$ values. However, the GA achieves a marginally higher result, with a $0.589\%$ increase in $|\mathscr{H}_i|$. This difference arises from the broader sampling of the parameter space inherent to the GA approach.

\subsection{Edge-Free Twisted Ring Cavity}~\label{sec:EF-Tw-RING-WG}

The EF twisted linear cavity described previously was further examined in a ring configuration, created by bending the underlying waveguide into a continuous closed loop. The total twist angle was constrained to discrete values of $\theta = 2\pi m$, $m \in \{0, 1, \dots, 8\}$, ensuring that the waveguide’s end faces joined seamlessly. 

For this geometry, the tetrahedral mesh parameters were set to a maximum element size of $4.89~\mathrm{mm}$ and a minimum element size of $0.209~\mathrm{mm}$. A maximum element growth rate of $1.35$ was used, together with a curvature factor of $0.3$. The narrow-region resolution was set to $0.85$.

The optimisation was performed using the GA framework. For each candidate geometry, an eigenvalue analysis was carried out to compute $35$ eigenfrequencies in the vicinity of $10$~GHz in order to optimise for the largest possible $|\mathscr{H}_i|$. 

\begin{figure}[t]
     \begin{center}
            \includegraphics[width=0.45\textwidth]{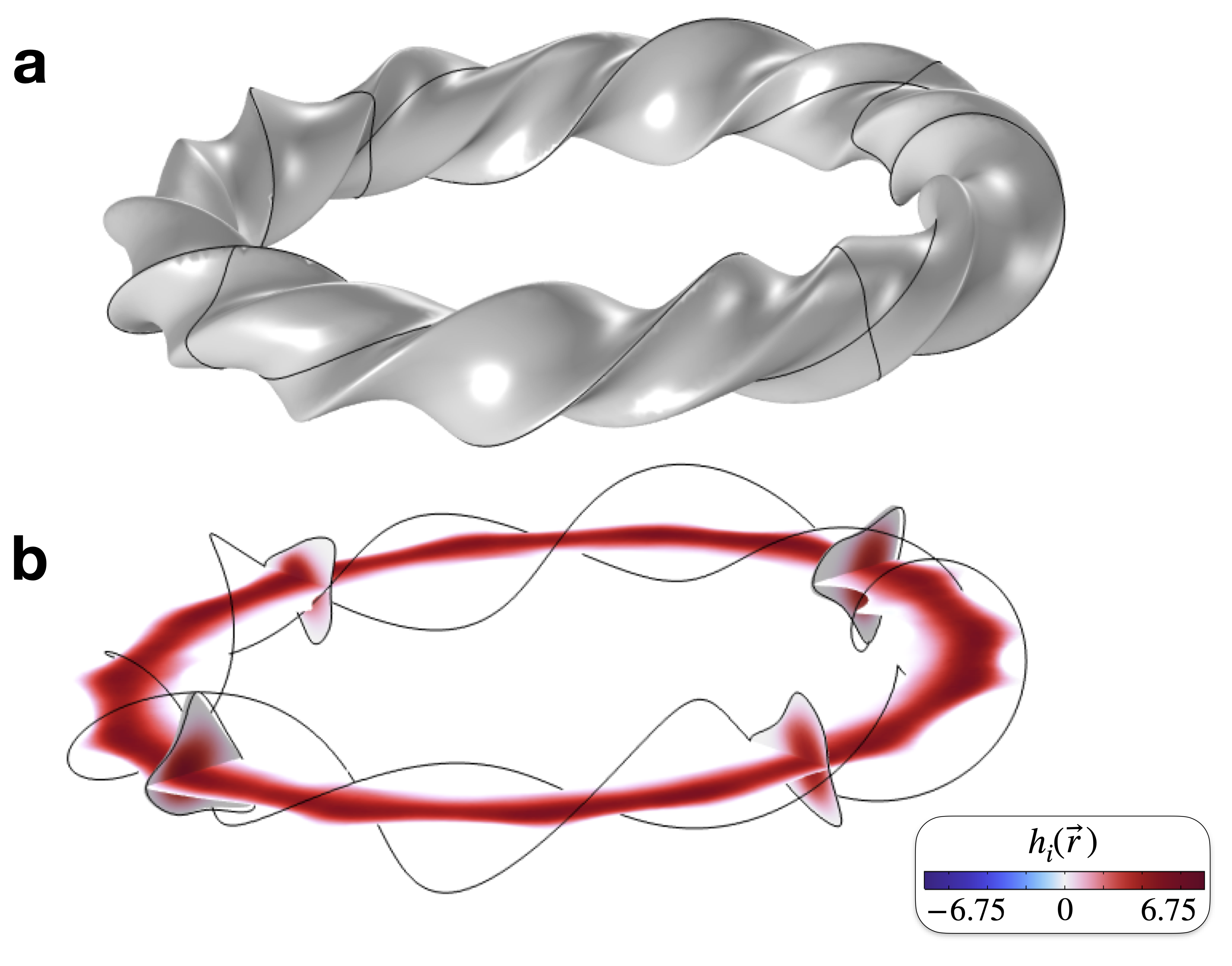}
            \end{center}
    \caption{(A) The optimised EF twisted ring cavity and (B) its $h_i(\vec{r})$ field distribution.}
   \label{fig:RING-Plots}
\end{figure}

The optimisation identified the configuration in Fig.~\ref{fig:RING-Plots}(a) as the highest-performing design, with the corresponding $h_i(\vec{r})$ field depicted in Fig.~\ref{fig:RING-Plots}(b). The geometric parameters associated with this optimised cavity are shown in Table~\ref{tab:optimised_linear_cavity}. The cavity achieves a very high $|\mathscr{H}_i|$ of $1.47$ and $\tilde{G}_i$ of $14.2~\mathrm{mm}$. 

The statistical robustness metrics for the GA-optimised EF twisted ring cavity under Gaussian geometric perturbations are summarised in Table~\ref{tab:robustness_metrics}. The very high $\bar{F}$ and $\mathcal{R}$ values indicate that the EF twisted ring cavity maintains large $|\mathscr{H}_i|$ even under geometric perturbations.

\subsection{Orthogonal Prism-intersection Cavity}~\label{sec:Cross_Cyl}

The orthogonal prism-intersection cavity geometry is generated by taking the volumetric intersection of two effectively infinite, untwisted waveguides whose cross-sectional boundaries are defined by closed $n$-point interpolation curves ($n=8$). Each curve lies in a transverse plane and is parameterised by radial distances $r_i$ (for the first prism) or $v_i$ (for the second) at fixed azimuthal angles $\varphi_i = 2\pi i/n$, consistent with the definition used in the previous section. Both prisms are assigned a length ($H=100~\mathrm{mm}$) much greater than their transverse dimensions so that end-cap effects become negligible. The cavity is then defined by the resulting intersection volume, which is fully determined by the two sets of radii $\{r_i\}$ and $\{v_i\}$. The free parameters defining the design space explored by the GA are the individual radii comprising these sets, each subject to the bounds ${r_i} \in [13, 33]~\mathrm{mm}$ and ${v_i} \in [13, 33]~\mathrm{mm}$. 

For this geometry, the tetrahedral mesh parameters were set to a maximum element size of $1.82~\mathrm{mm}$ and a minimum element size of $0.132~\mathrm{mm}$. A maximum element growth rate of $1.4$ was used, together with a curvature factor of $0.4$. The narrow-region resolution was set to $0.7$. 

The optimisation was performed using the GA framework. For each candidate geometry, $44$ eigenfrequencies near $10$ GHz were computed to evaluate the $|\mathscr{H}_i|$ of the supported eigenmodes.

\begin{figure}[t]
     \begin{center}
            \includegraphics[width=0.45\textwidth]{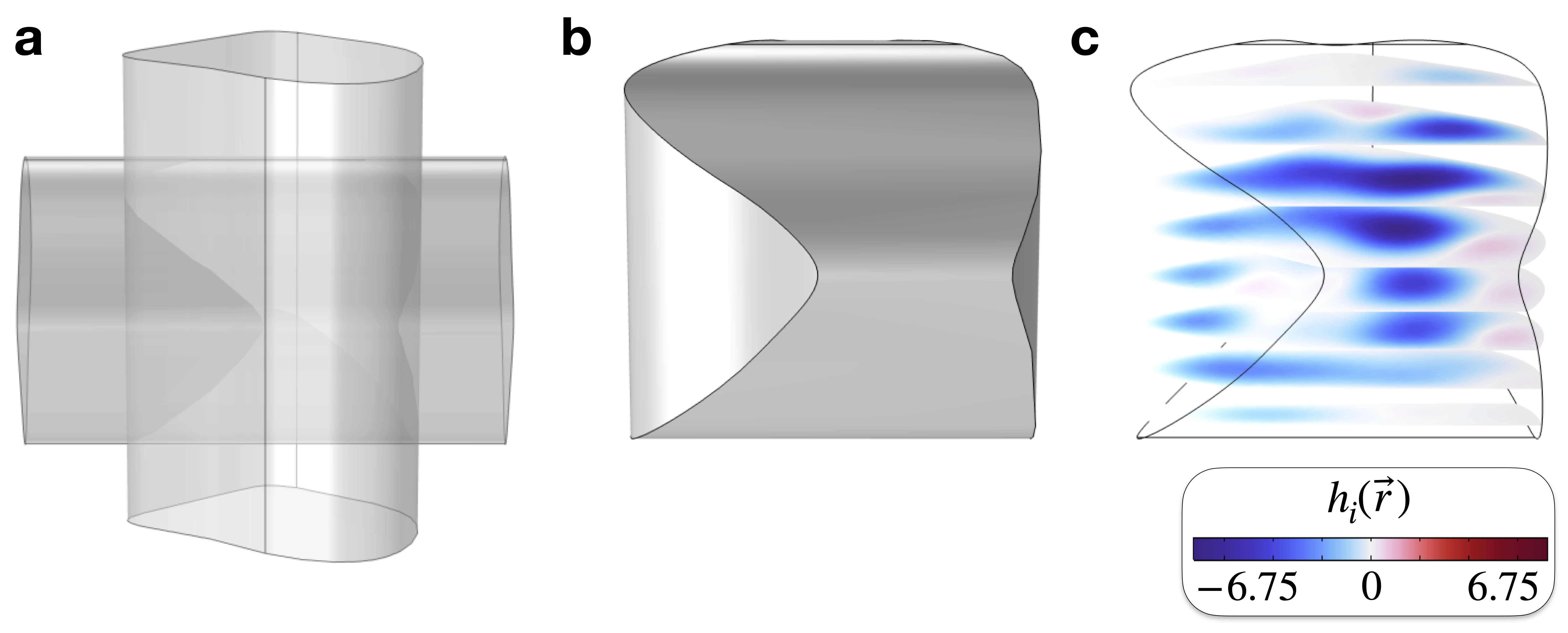}
            \end{center}
    \caption{(A) Two infinitely long, untwisted waveguides whose cross-sectional boundaries define (B) the optimised orthogonal prism-intersection cavity. (C) The corresponding $h_i(\vec{r})$ field distribution.}
   \label{fig:INTERSECTION-Plots}
\end{figure}

The construction process and resulting geometry of the optimised orthogonal prism-intersection cavity are illustrated in Fig.~\ref{fig:INTERSECTION-Plots}(a) and (b), with the $h_i(\vec{r})$ distribution of the resonant mode shown in Fig.~\ref{fig:INTERSECTION-Plots}(c). The corresponding optimised cavity parameters are $\{r_i\}=$$\{19.6, 15.9, 28.9, 18.3, 21.8, 21.9, 31.1, 32.5\}~\mathrm{mm}$ and $\{v_i\}=$$\{33.0, 19.2, 19.0, 13.5,20.8, 31.5, 30.7, 22.2\}~\mathrm{mm}$. This design achieved a $\mathscr{H}_i = -0.9$ and a $\tilde{G}_i = 8.27~\mathrm{mm}$.

The statistical robustness metrics for the GA-optimised orthogonal prism-intersection resonator under Gaussian geometric perturbations are summarised in Table~\ref{tab:robustness_metrics}. On average, the cavity maintained a high $|\mathscr{H}_i|$, though it exhibits only moderate robustness to geometric perturbations. This sensitivity arises because even small changes to the underlying cylindrical shapes significantly alter the intersection boundary that defines the cavity, which in turn strongly affects the helical field pattern.

\begin{table*}[t]
\centering
\caption{Optimised parameters for the sphere-subtracted cylindrical resonator obtained using GA and BO.}
\label{tab:optimised_sculpture_resonator}
\setlength{\tabcolsep}{5pt} 
\renewcommand{\arraystretch}{1.1}
\begin{tabular}{cccccc}
Optimisation & $H$ & $R$ & $\{\phi_i\}$ & $\{r_i\}$ (mm) & $\{dr_i\}$ (mm) \\
\hline
GA &
41.0 & 20.7 & \makecell[l]{\{1.08, 1.80, 3.21, 4.74, \\ 2.66, 5.28, 4.16, 3.22\}} & \makecell[l]{\{29.9, 36.1, 49.9, 30.5, \\ 47.1, 34.0, 49.1, 53.6\}} & \makecell[l]{\{5.42, 20.1, 19.2, 23.3, \\ 23.3, 11.9, 18.3, 24.3\}} \\[4pt]
BO & 39.2 & 47.0 & \makecell[l]{\{4.76, 3.22, 0.609, 3.77, \\ 0.756, 4.32, 2.65, 3.89\}} & \makecell[l]{\{54.7, 31.0, 45.6, 41.6, \\ 50.0, 54.7, 51.6, 23.0\}} & \makecell[l]{\{20.2, 12.7, 28.4, 14.2, \\ 22.3, 12.5, 17.3, 23.3\}} \\
\end{tabular}
\end{table*}

\begin{figure}[t]
     \begin{center}
            \includegraphics[width=0.45\textwidth]{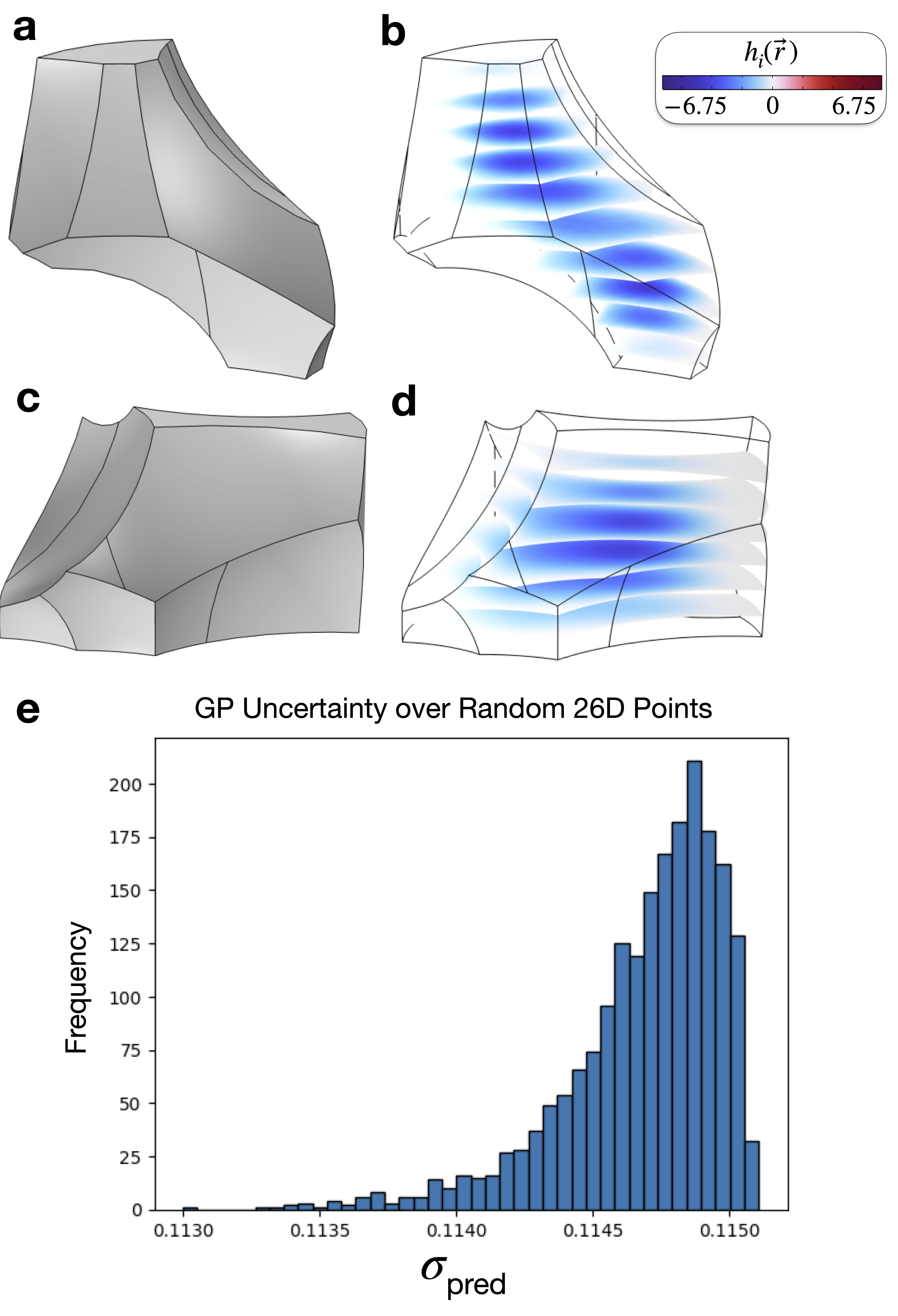}
            \end{center}
    \caption{(A) GA-optimised cavity based on sphere-subtracted sculpted cylinders and (B) its corresponding $h_i(\vec{r})$ field distribution. (C) BO-optimised cavity and (D) its corresponding $h_i(\vec{r})$ field distribution. (E) Histogram of $\sigma_{\text{pred}}$ across 2,000 random points in the 26D input space.}
   \label{fig:CURVING-LARGE-Plots}
\end{figure}

\subsection{Sphere-Subtracted Cylindrical Cavity}~\label{sec:sculpture_cavity}

A baseline cylindrical resonator of height $H$ and radius $R$ is first defined, with both treated as optimisation variables. Around the cylinder's top and bottom rims, a set of $n$-spherical voids is introduced ($n=8$), each described by its radius, $r_i$, radial offset from the central axis, $dr_i$, and azimuthal angle, $\phi_i$. The centres of spheres $i = 1$-$4$ are positioned a distance $H_1=10~\mathrm{mm}$ above the top rim, while those of spheres $i = 5$-$8$ are placed a distance $H_1$ below the bottom rim. The cavity geometry is then obtained by subtracting the union of these spheres from the solid cylinder. The design space for this cavity family is therefore defined by the free parameters, along with their respective ranges, $H\in [40,70]~\text{mm}$, $R\in [20,40]~\text{mm}$, $\phi_i \in [0,2\pi]$, $r_i \in [20,55]~\text{mm}$, and $dr_i \in [5,30]~\text{mm}$. 

The tetrahedral mesh used for this geometry employed a maximum element size of $3~\mathrm{mm}$ and a minimum element size of $0.2~\mathrm{mm}$. A maximum element growth rate of $1.4$ was used, together with a curvature factor of $0.4$. The narrow-region resolution was set to $0.7$.

The optimisation was performed using both the GA and BO framework. For each candidate geometry, $41$ eigenfrequencies near $5$ GHz were computed to evaluate the corresponding $|\mathscr{H}_i|$ values. 

The GA-optimised configuration is shown in Fig.~\ref{fig:CURVING-LARGE-Plots}(a), with the corresponding $h_i(\vec{r})$ field distribution depicted in Fig.~\ref{fig:CURVING-LARGE-Plots}(b). This design achieved a $\mathscr{H}_i$ of $-0.958$ and a $\tilde{G}_i$ of $3.83~\mathrm{mm}$. In comparison, BO applied to the same parameter space yielded an optimised resonator with $\mathscr{H}_i = -0.750$ and $\tilde{G}_i = 7.38~\mathrm{mm}$. The geometry of this resonator is shown in Fig.~\ref{fig:CURVING-LARGE-Plots}(c), and the corresponding $h_i(\vec{r})$ field is shown in Fig.~\ref{fig:CURVING-LARGE-Plots}(d). The parameters of these optimised geometries are summarised in Table~\ref{tab:optimised_sculpture_resonator}. 

While both optimisation methods produce large $|\mathscr{H}_i|$, the GA-optimised geometry achieves a $27.7\%$ higher value, whereas its $\tilde{G}_i$ is only half that of the BO-optimised design. This difference in $|\mathscr{H}_i|$ and $\tilde{G}_i$ arises from their differing search strategies: GA samples the true, potentially rugged fitness landscape and can access narrow, high-variance regions where strong field hybridisation and sharper curvature enhance $|\mathscr{H}_i|$ but also increase surface area, concentrating magnetic-field intensity near the conducting walls and lowering $\tilde{G}_i$. In contrast, BO relies on a smooth Gaussian-process surrogate that favours broad, slowly varying regions of parameter space, yielding geometries with gentler curvature.

The statistical robustness metrics for the GA-optimised sphere-subtracted cylindrical resonator under Gaussian geometric perturbations are summarised in Table~\ref{tab:robustness_metrics}. The $\bar{F}$ and $\mathcal{R}$ values indicate that the sphere-subtracted cylindrical resonator maintains high $|\mathscr{H}_i|$ on average but exhibits moderate sensitivity to geometric perturbations. This sensitivity to geometric perturbations likely arises from sharper geometric transitions inherent to the sphere-subtracted cylindrical design, which amplify local field variations under perturbation. 

For the BO-optimised surrogate, predictive uncertainty was evaluated over 2,000 random samples in the 26D design space. As shown in Fig.~\ref{fig:CURVING-LARGE-Plots}(e), the predicted standard deviations are tightly clustered between 0.114 and 0.115, with most values near the upper end of this narrow range. This near-uniformity confirms that the GP model is well-calibrated and confident across the explored region. 

\subsection{Parameterised Surface Cavity}~\label{sec:param_surface}

Two parametric surfaces, separated by $z_1 = 0.5~\mathrm{mm}$, are defined by a dimensionless parameter domain $s_1, s_2 \in [-1,1]$. The surfaces are described by
\begin{align}
\vec{r}_1(s_1,s_2)
=&
\Big(
s_1,\,
s_2,\,
p_1 \cos(\pi s_1)\sin(\pi s_2)
+ p_2 s_1 s_2 e^{-s_1^2 - s_2^2}\notag\\
&+ p_3 \cos(\pi s_1 s_2)
\Big)\,\text{mm},
\\[4pt]
\vec{r}_2(s_1,s_2)
=&
\Big(
s_1,\,
s_2,\,
p_4 \cos(\pi s_1)\sin(\pi s_2)
+ p_5 s_1 s_2 e^{-s_1^2 - s_2^2}\notag\\
&+ p_6 \cos(\pi s_1 s_2)
\Big)\,\text{mm},
\end{align}
where each $p_i \in [-1,1]~\mathrm{mm}$ controls the out-of-plane deformation of the surface. The cavity geometry is defined as the solid volume of a cylinder of radius $1~\mathrm{mm}$ whose upper and lower faces are replaced by the two parametric surfaces $\vec{r}_1$ and $\vec{r}_2$ over the domain $s_1, s_2 \in [-1,1]$. This volume is then uniformly scaled by a factor of 13 in all spatial dimensions. 

\begin{figure}[t]
     \begin{center}
            \includegraphics[width=0.45\textwidth]{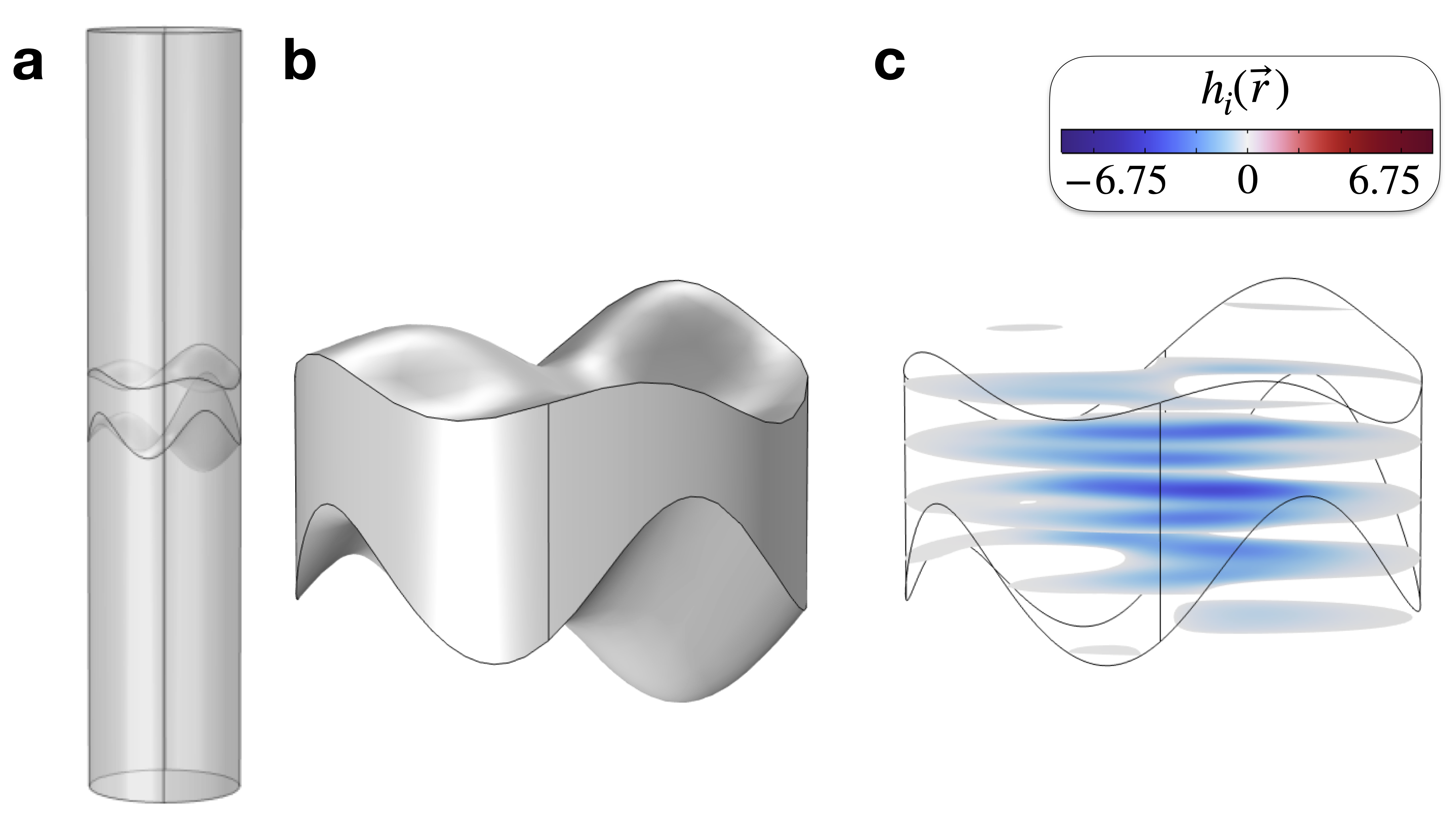}
            \end{center}
    \caption{(A) The original unmodified parameterised surfaces and cylinder, alongside (B) the optimised cavity formed from parameterised surfaces, and (C) the corresponding $h_i(\vec{r})$ field distribution.}
   \label{fig:PARAMETRISED-SURFACE-Plots}
\end{figure}

\begin{table*}[t]
\centering
\caption{Summary of the performance metrics of the optimised cavity geometries, showing the mode frequency $f_i$, electromagnetic helicity $\mathscr{H}_i$, frequency-normalised geometric factor $\tilde{G}_i$, and robustness index $\mathcal{R}$. The cylindrical cavity resonators use the same geometric dimensions as those reported in Ref.~\cite{Creedon2016}, with $f_i$ and $G_i$ computed independently and in agreement with the published values within numerical tolerance. The linear twisted $D_3$ resonators correspond to the geometry studied in Refs.~\cite{Bourhill_twisted_anyon_cavity_2023,paterson2025electromagnetic}, evaluated here at a twist angle of $\theta = 10\pi/9$.}
\label{tab:optimised_geometry_helicity_summary}
\begin{tabular}{l c c c c c}
Cavity & Optimisation & $f_i$ (GHz) & $\mathscr{H}_i$ &
$\tilde{G}_i$ (mm) & $\mathcal{R}$ \\
\hline
EF twisted linear (baseline) [Sec.~\ref{sec:EF-Tw-WG}] & GA &  8.57 &  1.03   &  5.26 & 1.02  \\
             & BO &  9.15 &  1.02   &  4.57 & ---   \\
EF twisted linear (expanded $r_i$)   & GA &  9.42 &  1.10   &  6.32 & 0.635 \\
EF twisted ring [Sec.~\ref{sec:EF-Tw-RING-WG}] & GA &  9.99 &  1.47   & 14.2  & 1.29  \\
Orthogonal prism-intersection [Sec.~\ref{sec:Cross_Cyl}] & GA & 12.0  & $-0.900$ &  8.27 & 0.735 \\
Sphere-subtracted cylindrical [Sec.~\ref{sec:sculpture_cavity}] & GA & 17.1  & $-0.958$ &  3.83 & 0.707 \\
                       & BO &  6.23 & $-0.750$ &  7.38 & ---   \\
Parametrised surface [Sec.~\ref{sec:param_surface}] & GA & 20.0  & $-0.673$ &  2.90 & 0.113 \\
Cylinder (TM$_{0,1,0}$)  & -- & 11.2  &  0      &  3.78 & ---   \\
Cylinder (TE$_{0,1,1}$)  & -- & 18.6  &  0      &  5.15 & ---   \\
Linear twisted $D_3$   & -- & 16.9  &  0.925  &  2.16 & ---   \\
\end{tabular}
\end{table*}

The tetrahedral mesh used for this configuration employed a maximum element size of $0.91~\mathrm{mm}$ and a minimum element size of $0.039~\mathrm{mm}$. A maximum element growth rate of $1.35$ was used, together with a curvature factor of $0.3$. The narrow-region resolution was set to $0.85$.

The optimisation was performed using the GA framework. For each candidate geometry, 44 eigenfrequencies in the vicinity of $10$ GHz were computed to evaluate $|\mathscr{H}_i|$. 

The surface-based construction method used to generate this optimised geometry is illustrated in Fig.~\ref{fig:PARAMETRISED-SURFACE-Plots}(a), the resulting structure is shown in Fig.~\ref{fig:PARAMETRISED-SURFACE-Plots}(b) and the corresponding $h_i(\vec{r})$ distribution is shown in Fig.~\ref{fig:PARAMETRISED-SURFACE-Plots}(c). The optimised geometry corresponds to the parameter set $\{p_i\} = $$\{-0.384, -0.864, -0.381, $$-0.192, -0.808, 0\}~\mathrm{mm}$. This cavity achieves $\mathscr{H}_i = -0.673$ and $\tilde{G}_i = 2.90~\mathrm{mm}$, indicating that this cavity family exhibits intrinsically low $|\mathscr{H}_i|$ and cannot be optimised toward the upper bound of $|\mathscr{H}_i|$. 

The statistical robustness metrics for the GA-optimised parameterised surface under Gaussian geometric perturbations are summarised in Table~\ref{tab:robustness_metrics}. The low $\mathcal{R}$ value indicates that the performance of the parameterised surface deteriorates substantially under small geometric deviations. This sensitivity to small geometric deviations arises from its highly variable surface topology, where even minor parameter shifts induce significant field distortions and reduce the stability of the chiral response.

\section{Discussion}

Inverse design enabled the discovery of fully 3D microwave cavities that support highly chiral eigenmodes, achieved by directly optimising cavity boundaries to maximise $|\mathscr{H}_i|$. This approach uncovers resonator geometries for which the required combinations of boundary-shape parameters would be difficult to anticipate from conventional, intuition-driven design methods and, in several cases, exhibits surface-loss sensitivity comparable to or better than that of standard cavity designs. All optimised cavity geometries are summarised in Table~\ref{tab:optimised_geometry_helicity_summary}, which also includes benchmark cylindrical cavity modes, TM$_{0,1,0}$ and TE$_{0,1,1}$, corresponding to the TM and TE modes of a cavity with an equal surface-to-volume ratio, as well as the fundamental helical mode of the twisted linear $D_3$ resonator with a twist angle of $\theta = 10\pi/9$~\cite{Bourhill_twisted_anyon_cavity_2023,paterson2025electromagnetic}.

Among all designs, the EF twisted resonator family consistently delivers the highest $|\mathscr{H}_i|$, owing to their global, continuous twist, which maintains a uniform spatial overlap of parallel components of $\vec{\mathbf{E}}$ and $\vec{\mathbf{H}}$ throughout the volume. As a result, $h_i$ integrates constructively over the entire cavity. The ring configuration of this resonator family emerges as the top performer across all metrics. Bending the linear geometry into a closed loop eliminates the metallic endcaps which would otherwise disrupt the uniform spatial overlap of $\vec{\mathbf{E}}$ and $\vec{\mathbf{H}}$, producing a larger $|\mathscr{H}_i|$. Their removal also eliminates sharp reflection planes, enforcing continuous boundaries around the cavity. Perturbations therefore act globally rather than being concentrated at end surfaces, enhancing $\mathcal{R}$. Expanding the allowable range of control-point radii can further increase $|\mathscr{H}_i|$ for the EF twisted linear resonator, but this enhancement arises from large radial excursions rather than global twist, making such helical modes more sensitive to dimensional variation.

The EF twisted linear resonator with modest bounds on the admissible range of control-point radii $r_i$, however, remains a strong performer and is easier to fabricate, as it avoids the need for seamless joining. While its $\tilde{G}_i$ is only $1.39$ times that of the TM$_{0,1,0}$ mode in a cylindrical cavity, the ring resonator achieves a substantially higher $\tilde{G}_i$ of $3.76$ times that of the TM$_{0,1,0}$ mode. This increase also arises from the elimination of the metallic endcaps, which reduces magnetic field confinement near the conducting walls and therefore lowers surface loss. 

To provide an estimate of achievable performance, we evaluate the corresponding intrinsic quality factors for realistic material parameters at microwave frequencies and millikelvin temperatures. For state-of-the-art superconducting niobium, surface resistances on the order of $R_S \sim 10^{-9}~\Omega$ have been reported, which would correspond to quality factors approaching $Q_0 \sim 10^{12}$ for the EF twisted ring resonator. For more readily accessible implementations, additively manufactured aluminium cavities exhibit surface resistances on the order of $R_S \sim 10^{-4}~\Omega$, implying that initial realisations of these designs could achieve $Q_0 \sim 10^{6}$. 

The orthogonal prism-intersection resonator is another high-performing design, exhibiting a $\tilde{G}_i$ higher than that of the EF twisted linear resonator and approximately $2.19$ times that of the TM$_{0,1,0}$ mode. Its favourable $\tilde{G}_i$ arises from the absence of deep recesses or sharply varying curvature that could pull the magnetic field towards the conducting walls, allowing for reduced sensitivity to surface losses. This resonator also supports a strong $|\mathscr{H}_i|$ and moderate $\mathcal{R}$. The moderate sensitivity to geometric variations likely stems from the fact that the $|\mathscr{H}_i|$ is generated by local curvature variations rather than a global twist; such locally induced features are more susceptible to perturbations. 

By contrast, the parameterised-surface resonator performs the worst across all metrics. Its deep, irregular curvature disrupts constructive integration of $h_i$ throughout the volume, suppressing $|\mathscr{H}_i|$, and draws the magnetic field toward the conducting boundaries, reducing $\tilde{G}_i$.

The sphere-subtracted cylindrical resonator exhibits moderately high $|\mathscr{H}_i|$, but a low $G_i$ and only moderate $\mathcal{R}$. In this case, the $|\mathscr{H}_i|$ arises from sharp, highly varying complex features and recesses that concentrate the magnetic field near the conducting walls, lowering $G_i$. Because the mode is supported by these intricate local features, it is also more susceptible to small geometric deviations, which reduce $\mathcal{R}$.

The inverse-design optimisation reveals that highly helical modes in these cavity families arise through two distinct mechanisms. The first mechanism arises in complex geometries containing sharp and/or rapidly varying curvature. In these designs, strong $|\mathscr{H}_i|$ is produced through the local curvature of these features. However, such $|\mathscr{H}_i|$ is inherently fragile: even small geometric deviations disrupt the localised spatial overlap of $\vec{\mathbf{E}}$ and $\vec{\mathbf{H}}$, leading to a reduced $\mathcal{R}$. As a result, although these geometries can exhibit high $|\mathscr{H}_i|$, they do so at the expense of $\mathcal{R}$. Their $\tilde{G}_i$ is sensitive to the depth of these features. Deep recesses pull a larger fraction of the magnetic field toward the conducting walls, increasing surface losses that lower $\tilde{G}_i$ as well as amplifying sensitivity to geometric perturbations. 

The second mechanism occurs in geometries with a global, continuous twist. Here, the $|\mathscr{H}_i|$ arises from a smooth global curvature that maintains a consistent spatial overlap of $\vec{\mathbf{E}}$ and $\vec{\mathbf{H}}$ along the entire structure. Therefore, $h_i$ integrates constructively throughout the full cavity volume, producing high $|\mathscr{H}_i|$ without relying on delicate local features. Because the $|\mathscr{H}_i|$ is reinforced globally rather than locally, these designs are far less sensitive to dimensional variations and therefore exhibit a superior $\mathcal{R}$. The strongest performers across all metrics are those with continuous boundaries and no sharp reflection planes or discontinuities. These results highlight the advantage of global twist over highly complex geometries when $\mathcal{R}$ is a key design requirement.

Comparing the performance of the GA- and BO-optimised cavities in Table~\ref{tab:optimised_geometry_helicity_summary}, it is evident that the GA identified solutions with higher $|\mathscr{H}_i|$ than BO, as also noted in Sec.~\ref{sec:sculpture_cavity}. This demonstrates the effectiveness of broad, stochastic search strategies when navigating complex 3D parameter spaces.

\section{Conclusion}

In this work, we demonstrated an inverse-design framework for engineering microwave cavity resonators with large electromagnetic helicity, $|\mathscr{H}_i|$. By formulating $|\mathscr{H}_i|$ maximisation as a boundary-shape optimisation problem and exploring the resulting design space using gradient-free optimisation strategies, the framework provides a systematic means of exploring the prescribed boundary-shape parameter space within each cavity family. This approach enables the identification of resonator geometries composed of smooth, edge-free components that support strongly helical modes, including optimised configurations with $|\mathscr{H}_i|$ exceeding that of the previously studied heuristically designed twisted cavities with $D_3$ symmetry~\cite{paterson2025electromagnetic,Bourhill_twisted_anyon_cavity_2023,paterson2025dynamicallytuneablehelicitytwisted,paterson_Berry}. 

The optimised 3D microwave resonators are also expected to exhibit favourable geometric surface-loss characteristics and strong robustness to geometric perturbations, with the robustness analysis indicating resilience to fabrication-induced imperfections such as those associated with additive manufacturing. In contrast to the broader inverse-design literature on photonic structures and microwave components, the framework developed here is specifically formulated for fully 3D metallic microwave cavity resonators and evaluates cavity eigenmodes within a defined broad microwave frequency window.

This same framework can be adapted to alternative or composite figures of merit, such as gradients of fields or joint optimisation of $|\mathscr{H}_i|$ and $\tilde{G}_i$, providing a flexible path toward tailored chiral-field engineering. The inverse-design approach therefore provides a practical framework developed for fully three-dimensional metallic microwave cavity resonators, with an eigenmode-agnostic optimisation that allows the mode maximising the FoM to change in both modal identity and field structure as the cavity geometry evolves. The resulting framework enables the systematic optimisation of complex 3D metallic boundaries designed with manufacturability in mind and can be tailored to applications in enantioselective spectroscopy, axion haloscopes, and beyond.

\section*{Acknowledgment}

This work was funded by the Australian Research Council Centre of Excellence for Engineered Quantum Systems (CE170100009) and the Centre of Excellence for Dark Matter Particle Physics (CE200100008). E.C.P. is partially funded by the Defence Science Centre, Department of the Premier and Cabinet (WA Government), through the Research Higher Degree Student Grant (Project No. 2526R7RHDSG017). We thank Michael Hatzon for proofreading the manuscript. 

\clearpage

\appendix[Heuristic Scaling of Genetic Algorithm Parameters]
\label{sec:GAHeuristics}

The genetic algorithm (GA) parameters used in this study were empirically tuned to ensure robust convergence across optimisation spaces of varying dimensionality $d$, implemented using the \texttt{PyGAD} Python Genetic Algorithm framework. The population size, $N_{\mathrm{pop}}$, was typically set between $(5 \ldots 10)\times d$ to maintain genetic diversity, while practical values ranged from $(2 \ldots 14)\times d$ depending on computational limits. The number of generations, $N_{\mathrm{gen}}$, was increased with $d$ to ensure convergence in higher-dimensional design spaces, generally spanning $50$ to $200$ generations. The number of parents, $N_{\mathrm{par}}$, determined how many of the top-performing individuals contributed to the next generation and was scaled with the population size to preserve crossover diversity. Elitism, $N_{\mathrm{elite}}$, specified the number of top individuals preserved unchanged between generations and was kept approximately constant as a fraction of the total population (typically 20--25\%), except where reduced to manage computational load. A summary of how these variables correspond to the underlying \texttt{PyGAD} parameter names is given in Table~\ref{tab:pygad_mapping}.

The mutation strength, $\sigma_{\mathrm{mut}}$, as defined in the main text (see Sec.~\ref{sec:methods}), typically ranged between 0.08--0.18 across cavity families to balance exploration and stability. The mutation rate, $f_{\mathrm{mut}}$, controlled the fraction of genes modified per generation and was set between 10--35\%, with higher values applied to geometries requiring broader exploration or exhibiting more complex optimisation landscapes. 

A \texttt{"single\_point"} crossover and \texttt{"tournament"} parent selection (size~3) were employed consistently across all optimisations, as these parameters primarily affect selection pressure rather than dimensionality. Convergence was monitored using a stopping criterion, $\mathcal{T}_{\mathrm{stop}}$, based on either fitness thresholds or saturation conditions (e.g., \texttt{"reach\_98"} or \texttt{"saturate\_7-40"}), ensuring termination once optimisation progress plateaued. 

\begin{table}[!t]
\caption{Mapping between the notation used in this work and the corresponding \texttt{PyGAD} parameters.}
\label{tab:pygad_mapping}
\begin{tabular}{c c}
\textbf{Manuscript Notation} & \textbf{PyGAD Parameter} \\
\hline
Design dimensionality \(d\)               & \texttt{num\_genes} \\
Population size \(N_{\mathrm{pop}}\)      & \texttt{sol\_per\_pop} \\
Number of generations \(N_{\mathrm{gen}}\) & \texttt{num\_generations} \\
Number of parents \(N_{\mathrm{par}}\) & \texttt{num\_parents\_mating} \\
Elitism count \(N_{\mathrm{elite}}\)       & \texttt{keep\_elitism} \\
Mutation rate \(f_{\mathrm{mut}}\)         & \texttt{mutation\_percent\_genes}/100 \\
Stopping criterion \(\mathcal{T}_{\mathrm{stop}}\) & \texttt{stop\_criteria} \\
\end{tabular}
\end{table}
   
\hspace{10pt}

\bibliographystyle{IEEEtran}
\bibliography{ML-1-IEEE-arxiv}

@article{gad2023pygad,
  title     = {{PyGAD}: An intuitive genetic algorithm {Python} library},
  author    = {Gad, Ahmed Fawzy},
  journal   = {Multimedia Tools and Applications},
  volume    = {83},
  number    = {20},
  pages     = {58029--58042},
  year      = {2024},
  publisher = {Springer},
  doi       = {10.1007/s11042-023-17167-y}
}

@article{Flisgen2006Eigenvalue,
  author  = {Herter, Christine and Schöps, Sebastian and Wollner, Winnifried},
  title   = {Eigenvalue Optimization with respect to Shape-Variations in Electromagnetic Cavities},
  journal = {Proceedings in Applied Mathematics and Mechanics},
  volume  = {22},
  number  = {1},
  pages   = {e202200122},
  doi     = {10.1002/pamm.202200122},
  url     = {https://onlinelibrary.wiley.com/doi/abs/10.1002/pamm.202200122},
  eprint  = {https://onlinelibrary.wiley.com/doi/pdf/10.1002/pamm.202200122},
  year    = {2023}
}

@article{Ziegler2024GradientBased,
  author   = {Ziegler, Anna and Hahn, Robert and Isensee, Victoria and Nguyen, Anh Duc and Schöps, Sebastian},
  title    = {Gradient-based eigenvalue optimization for electromagnetic cavities with built-in mode matching},
  journal  = {IET Science, Measurement \& Technology},
  volume   = {18},
  number   = {9},
  pages    = {462--470},
  doi      = {10.1049/smt2.12190},
  url      = {https://ietresearch.onlinelibrary.wiley.com/doi/abs/10.1049/smt2.12190},
  eprint   = {https://ietresearch.onlinelibrary.wiley.com/doi/pdf/10.1049/smt2.12190},
  year     = {2024}
}

@article{scikit-learn,
  title   = {Scikit-learn: Machine Learning in {P}ython},
  author  = {Pedregosa, F. and Varoquaux, G. and Gramfort, A. and Michel, V.
             and Thirion, B. and Grisel, O. and Blondel, M. and Prettenhofer, P.
             and Weiss, R. and Dubourg, V. and Vanderplas, J. and Passos, A. and
             Cournapeau, D. and Brucher, M. and Perrot, M. and Duchesnay, E.},
  journal = {Journal of Machine Learning Research},
  volume  = {12},
  pages   = {2825--2830},
  year    = {2011}
}

@article{paterson2025dynamicallytuneablehelicitytwisted,
  author  = {Paterson, E. C. I. and Bourhill, J. and Tobar, M. E. and Goryachev, M.},
  title   = {Dynamically tunable helicity in twisted electromagnetic resonators},
  journal = {Phys. Rev. A},
  volume  = {113},
  number  = {3},
  pages   = {033506},
  year    = {2026},
  month   = mar,
  doi     = {10.1103/zggj-w33f}
}

@article{Bliokh_2013,
  author    = {Konstantin Y Bliokh and Aleksandr Y Bekshaev and Franco Nori},
  journal   = {New Journal of Physics},
  pages     = {033026},
  publisher = {IOP Publishing},
  title     = {Dual electromagnetism: helicity, spin, momentum, and angular momentum (2013 New J. Phys. 15 033026)},
  url       = {https://iopscience.iop.org/article/10.1088/1367-2630/15/3/033026},
  volume    = {15},
  year      = {2015}
}

@article{Multunas_Circular_Dichroism_Crystals,
  author    = {Multunas, Christian and Grieder, Andrew and Xu, Junqing and Ping, Yuan and Sundararaman, Ravishankar},
  doi       = {10.1103/PhysRevMaterials.7.123801},
  issue     = {12},
  journal   = {Phys. Rev. Mater.},
  month     = dec,
  numpages  = {11},
  pages     = {123801},
  publisher = {American Physical Society},
  title     = {Circular dichroism of crystals from first principles},
  url       = {https://link.aps.org/doi/10.1103/PhysRevMaterials.7.123801},
  volume    = {7},
  year      = {2023}
}

@article{Hendry:2010ug,
  author  = {Hendry, E. and Carpy, T. and Johnston, J. and Popland, M. and Mikhaylovskiy, R. V. and Lapthorn, A. J. and Kelly, S. M. and Barron, L. D. and Gadegaard, N. and Kadodwala, M.},
  date    = {2010/11/01},
  doi     = {10.1038/nnano.2010.209},
  isbn    = {1748-3395},
  journal = {Nature Nanotechnology},
  number  = {11},
  pages   = {783--787},
  title   = {Ultrasensitive detection and characterization of biomolecules using superchiral fields},
  url     = {https://doi.org/10.1038/nnano.2010.209},
  volume  = {5},
  year    = {2010}
}

@article{Tang2011,
  author  = {{Tang}, Yiqiao and {Cohen}, Adam E.},
  title   = {{Enhanced Enantioselectivity in Excitation of Chiral Molecules by Superchiral Light}},
  journal = {Science},
  year    = 2011,
  month   = apr,
  volume  = {332},
  number  = {6027},
  pages   = {333},
  doi     = {10.1126/science.1202817}
}

@article{UpAndComing_Advances_Kutsaev_Advanced_Photonics_2021,
  author  = {Kutsaev, Sergey V. and Krasnok, Alex and Romanenko, Sergey N. and Smirnov, Alexander Yu. and Taletski, Kirill and Yakovlev, Vyacheslav P.},
  title   = {Up-And-Coming Advances in Optical and Microwave Nonreciprocity: From Classical to Quantum Realm},
  journal = {Advanced Photonics Research},
  volume  = {2},
  number  = {3},
  pages   = {2000104},
  doi     = {10.1002/adpr.202000104},
  url     = {https://advanced.onlinelibrary.wiley.com/doi/abs/10.1002/adpr.202000104},
  year    = {2021}
}

@article{Both2022,
  author  = {Both, Steffen and Sch\"aferling, Martin and Sterl, Florian and Muljarov, Egor A. and Giessen, Harald and Weiss, Thomas},
  title   = {Nanophotonic Chiral Sensing: How Does It Actually Work?},
  journal = {ACS Nano},
  volume  = {16},
  number  = {2},
  pages   = {2822--2832},
  year    = {2022},
  month   = jan,
  issn    = {1936-0851},
  doi     = {10.1021/acsnano.1c09796},
  url     = {https://doi.org/10.1021/acsnano.1c09796},
  eprint  = {https://pubs.acs.org/ancac3/article-pdf/16/2/2822/15478349/nn1c09796.pdf}
}

@article{Toward_quantum_sensing_Volker_2023,
  author  = {V{\"o}lker, Laura A. and Herb, Konstantin and Janitz, Erika and Degen, Christian L. and Abendroth, John M.},
  title   = {Toward quantum sensing of chiral induced spin selectivity: Probing donor--bridge--acceptor molecules with {NV} centers in diamond},
  journal = {The Journal of Chemical Physics},
  volume  = {158},
  number  = {16},
  pages   = {161103},
  year    = {2023},
  doi     = {10.1063/5.0145466}
}

@article{Wang2021Intelligent,
  author  = {Wang, Ning and Yan, Wei and Qu, Yuhui and Li, Siyuan and Liu, Mengxin and Qiu, Min and Liu, Zhaowei},
  title   = {Intelligent Designs in Nanophotonics: From Optimization Towards Inverse Creation},
  journal = {PhotoniX},
  year    = {2021},
  volume  = {2},
  pages   = {22},
  doi     = {10.1186/s43074-021-00044-y}
}

@article{Kim2025,
  author  = {Kim, J. and Kim, J. Y. and Kim, J. and Hyeong, Y. and Neseli, B. and You, J. B. and Shim, J. and Shin, J. and Park, H. H. and Kurt, H.},
  title   = {Inverse Design of Nanophotonic Devices Enabled by Optimization Algorithms and Deep Learning: Recent Achievements and Future Prospects},
  journal = {Nanophotonics},
  year    = {2025},
  volume  = {14},
  number  = {2},
  pages   = {121--151},
  doi     = {10.1515/nanoph-2024-0536},
  pmcid   = {PMC11806510}
}

@article{Kang2024LargeScale,
  author  = {Kang, Chanik and Park, Chaejin and Lee, Myunghoo and Kang, Joonho and Jang, Min Seok and Chung, Haejun},
  title   = {Large-Scale Photonic Inverse Design: Computational Challenges and Breakthroughs},
  journal = {Nanophotonics},
  volume  = {13},
  number  = {20},
  pages   = {3765--3792},
  year    = {2024},
  doi     = {10.1515/nanoph-2024-0127}
}

@article{Wiecha2022,
  author  = {{Wiecha}, Peter R. and {Petrov}, Alexander Yu. and {Genevet}, Patrice and {Bogdanov}, Andrey},
  title   = {{Inverse design of nanophotonics devices and materials}},
  journal = {Photonics and Nanostructures},
  year    = 2022,
  month   = dec,
  volume  = {52},
  eid     = {101084},
  pages   = {101084},
  doi     = {10.1016/j.photonics.2022.101084},
  adsurl  = {https://ui.adsabs.harvard.edu/abs/2022PhNan..5201084W}
}

@book{Deng2022,
  author    = {Y. Deng},
  title     = {Adjoint Topology Optimization Theory for Nano-Optics},
  publisher = {Springer Nature Singapore},
  address   = {Singapore},
  year      = {2022},
  doi       = {10.1007/978-981-16-7969-8}
}

@article{Gedeon2023,
  author  = {J. Gedeon and E. Hassan and A. Cal{\`a} Lesina},
  title   = {Time-Domain Topology Optimization of Arbitrary Dispersive Materials for Broadband 3-D Nanophotonics Inverse Design},
  journal = {ACS Photonics},
  volume  = {10},
  number  = {11},
  pages   = {3875--3887},
  month   = nov,
  year    = {2023},
  doi     = {10.1021/acsphotonics.3c00572}
}

@article{Hammond2022,
  author  = {A. M. Hammond and A. Oskooi and M. Chen and Z. Lin and S. G. Johnson and S. E. Ralph},
  title   = {High-Performance Hybrid Time/Frequency-Domain Topology Optimization for Large-Scale Photonics Inverse Design},
  journal = {Opt. Express},
  volume  = {30},
  number  = {3},
  pages   = {4467--4491},
  month   = jan,
  year    = {2022},
  doi     = {10.1364/OE.442074}
}

@article{Hughes_2018,
  author  = {T. W. Hughes and M. Minkov and I. A. D. Williamson and S. Fan},
  title   = {Adjoint Method and Inverse Design for Nonlinear Nanophotonic Devices},
  journal = {ACS Photonics},
  volume  = {5},
  number  = {12},
  pages   = {4781--4787},
  month   = dec,
  year    = {2018},
  doi     = {10.1021/acsphotonics.8b01522}
}

@article{Schul2026,
  author  = {Schul, Louise and Mason, Sydney and Eun, Sungjun and Ahn, Geun Ho and Vučković, Jelena},
  title   = {Inverse Design for Scalable Photonic Systems},
  journal = {Nature Reviews Materials},
  year    = {2026},
  doi     = {10.1038/s41578-026-00915-5}
}

@article{Molesky_2018,
  author  = {Molesky, Sean and Lin, Zin and Piggott, Alexander Y. and Jin, Weiliang and Vuckovi{\'c}, Jelena and Rodriguez, Alejandro W.},
  title   = {Inverse Design in Nanophotonics},
  journal = {Nature Photonics},
  year    = {2018},
  volume  = {12},
  number  = {11},
  pages   = {659--670},
  doi     = {10.1038/s41566-018-0246-9},
  url     = {https://doi.org/10.1038/s41566-018-0246-9},
  issn    = {1749-4893}
}

@article{Invisible_Axion_Search_2021_Sikivie,
  title     = {Invisible axion search methods},
  author    = {Sikivie, Pierre},
  journal   = {Rev. Mod. Phys.},
  volume    = {93},
  issue     = {1},
  pages     = {015004},
  numpages  = {36},
  year      = {2021},
  month     = feb,
  publisher = {American Physical Society},
  doi       = {10.1103/RevModPhys.93.015004},
  url       = {https://link.aps.org/doi/10.1103/RevModPhys.93.015004}
}

@article{Bourhill_twisted_anyon_cavity_2023,
  author    = {Bourhill, J. F. and Paterson, E. C. I. and Goryachev, M. and Tobar, M. E.},
  doi       = {10.1103/PhysRevD.108.052014},
  issue     = {5},
  journal   = {Phys. Rev. D},
  month     = sep,
  numpages  = {11},
  pages     = {052014},
  publisher = {American Physical Society},
  title     = {Searching for ultralight axions with twisted cavity resonators of anyon rotational symmetry with bulk modes of nonzero helicity},
  url       = {https://link.aps.org/doi/10.1103/PhysRevD.108.052014},
  volume    = {108},
  year      = {2023}
}

@article{PhysRevLett.113.033601,
  author    = {Bliokh, Konstantin Y. and Kivshar, Yuri S. and Nori, Franco},
  doi       = {10.1103/PhysRevLett.113.033601},
  issue     = {3},
  journal   = {Phys. Rev. Lett.},
  month     = jul,
  numpages  = {6},
  pages     = {033601},
  publisher = {American Physical Society},
  title     = {Magnetoelectric Effects in Local Light-Matter Interactions},
  url       = {https://link.aps.org/doi/10.1103/PhysRevLett.113.033601},
  volume    = {113},
  year      = {2014}
}

@article{Alpeggiani_Electromagnetic_2018,
  author    = {Alpeggiani, F. and Bliokh, K. Y. and Nori, F. and Kuipers, L.},
  doi       = {10.1103/PhysRevLett.120.243605},
  issue     = {24},
  journal   = {Phys. Rev. Lett.},
  month     = jun,
  numpages  = {6},
  pages     = {243605},
  publisher = {American Physical Society},
  title     = {Electromagnetic Helicity in Complex Media},
  url       = {https://link.aps.org/doi/10.1103/PhysRevLett.120.243605},
  volume    = {120},
  year      = {2018}
}

@article{Melia_2020,
  author  = {Melia, Michael and Duran, Jesse and Koepke, Joshua and Saiz, David and Jared, Bradley and Schindelholz, Eric},
  year    = {2020},
  month   = dec,
  pages   = {},
  title   = {How build angle and post-processing impact roughness and corrosion of additively manufactured {316L} stainless steel},
  volume  = {4},
  journal = {npj Materials Degradation},
  doi     = {10.1038/s41529-020-00126-5}
}

@article{surface_roughness_Giovanni,
  title   = {Surface roughness analysis, modelling and prediction in selective laser melting},
  journal = {Journal of Materials Processing Technology},
  volume  = {213},
  number  = {4},
  pages   = {589--597},
  year    = {2013},
  issn    = {0924-0136},
  doi     = {10.1016/j.jmatprotec.2012.11.011},
  url     = {https://www.sciencedirect.com/science/article/pii/S0924013612003366},
  author  = {Giovanni Strano and Liang Hao and Richard M. Everson and Kenneth E. Evans}
}

@article{Microstructures_Rafi_2013,
  author  = {{Rafi}, H.~K. and {Karthik}, N.~V. and {Gong}, Haijun and {Starr}, Thomas L. and {Stucker}, Brent E.},
  title   = {{Microstructures and Mechanical Properties of Ti6Al4V Parts Fabricated by Selective Laser Melting and Electron Beam Melting}},
  journal = {Journal of Materials Engineering and Performance},
  year    = 2013,
  month   = dec,
  volume  = {22},
  number  = {12},
  pages   = {3872--3883},
  doi     = {10.1007/s11665-013-0658-0}
}

@article{morphological_Hamidi_2018,
  title   = {On morphological surface features of the parts printed by selective laser melting {SLM}},
  journal = {Additive Manufacturing},
  volume  = {24},
  pages   = {373--377},
  year    = {2018},
  issn    = {2214-8604},
  doi     = {10.1016/j.addma.2018.10.011},
  url     = {https://www.sciencedirect.com/science/article/pii/S2214860418304998},
  author  = {Milad Hamidi Nasab and Dario Gastaldi and Nora Francesca Lecis and Maurizio Vedani}
}

@article{Wi_2019_Electropolishing,
  author         = {Wu, Yao-Cheng and Kuo, Che-Nan and Chung, Yueh-Chun and Ng, Chee-How and Huang, Jacob C.},
  title          = {Effects of Electropolishing on Mechanical Properties and Bio-Corrosion of {Ti6Al4V} Fabricated by Electron Beam Melting Additive Manufacturing},
  journal        = {Materials},
  volume         = {12},
  year           = {2019},
  number         = {9},
  article-number = {1466},
  url            = {https://www.mdpi.com/1996-1944/12/9/1466},
  issn           = {1996-1944},
  doi            = {10.3390/ma12091466}
}

@inproceedings{Tyagi2018,
  author    = {Tyagi, P. and Goulet, T. and Brent, D. and Klein, K. and Garcia-Moreno, F.},
  title     = {Scanning Electron Microscopy and Optical Profilometry of Electropolished Additively Manufactured 316 Steel Components},
  booktitle = {Proc. ASME 2018 Int. Mech. Eng. Congr. Expo. (IMECE2018)},
  volume    = {2},
  year      = {2018},
  doi       = {10.1115/IMECE2018-88339}
}

@article{chaghazardi2022review,
  author  = {{Chaghazardi}, Z. and {W{\"u}thrich}, R.},
  title   = {{Review-Electropolishing of Additive Manufactured Metal Parts}},
  journal = {Journal of the Electrochemical Society},
  year    = 2022,
  month   = apr,
  volume  = {169},
  number  = {4},
  eid     = {043510},
  pages   = {043510},
  doi     = {10.1149/1945-7111/ac6450}
}

@article{paterson_Berry,
  author  = {E. C. I. Paterson and M. E. Tobar and M. Goryachev and J. Bourhill},
  title   = {Distinct Berry Phases in a Single Triangular M{\"o}bius Microwave Resonator},
  journal = {Phys. Rev. A},
  volume  = {113},
  number  = {4},
  pages   = {043501},
  month   = apr,
  year    = {2026},
  doi     = {10.1103/qnym-rzrs}
}

@article{paterson2025electromagnetic,
  title     = {Electromagnetic helicity in twisted cavity resonators},
  author    = {Paterson, E. C. I. and Bourhill, J. and Tobar, M. E. and Goryachev, M.},
  journal   = {Phys. Rev. A},
  volume    = {112},
  issue     = {1},
  pages     = {013530},
  numpages  = {12},
  year      = {2025},
  month     = jul,
  publisher = {American Physical Society},
  doi       = {10.1103/6gp4-76td},
  url       = {https://link.aps.org/doi/10.1103/6gp4-76td}
}

@article{LANDOLT19871,
  title   = {Fundamental aspects of electropolishing},
  journal = {Electrochimica Acta},
  volume  = {32},
  number  = {1},
  pages   = {1--11},
  year    = {1987},
  issn    = {0013-4686},
  doi     = {10.1016/0013-4686(87)87001-9},
  url     = {https://www.sciencedirect.com/science/article/pii/0013468687870019},
  author  = {D. Landolt}
}

@inbook{Aryan2015,
  author    = {Aryan, Naser Pour and Kaim, Hans and Rothermel, Albrecht},
  title     = {Primary Current Distribution and Electrode Geometry},
  booktitle = {Stimulation and Recording Electrodes for Neural Prostheses},
  publisher = {Springer International Publishing},
  address   = {Cham},
  pages     = {25--30},
  year      = {2015},
  doi       = {10.1007/978-3-319-10052-4_4}
}

@article{TYAGI201932,
  title   = {Reducing the roughness of internal surface of an additive manufacturing produced 316 steel component by chempolishing and electropolishing},
  journal = {Additive Manufacturing},
  volume  = {25},
  pages   = {32--38},
  year    = {2019},
  issn    = {2214-8604},
  doi     = {10.1016/j.addma.2018.11.001},
  url     = {https://www.sciencedirect.com/science/article/pii/S2214860418306079},
  author  = {Pawan Tyagi and Tobias Goulet and Christopher Riso and Robert Stephenson and Nitt Chuenprateep and Justin Schlitzer and Cordell Benton and Francisco Garcia-Moreno}
}

@article{PEREZ2015352,
  title   = {Simulation of current distribution along a planar electrode under turbulent flow conditions in a laboratory filter-press flow cell},
  journal = {Electrochimica Acta},
  volume  = {154},
  pages   = {352--360},
  year    = {2015},
  issn    = {0013-4686},
  doi     = {10.1016/j.electacta.2014.11.166},
  url     = {https://www.sciencedirect.com/science/article/pii/S0013468614023962},
  author  = {Tzayam Pérez and Carlos {Ponce de León} and Frank C. Walsh and José L. Nava}
}

@article{Taguchi2025,
  title     = {Nanoscale chirality enhancement using topology-designed three-dimensional dielectric nanogap antennas},
  author    = {Taguchi, Atsushi and Fukui, Yamato and Sasaki, Keiji},
  journal   = {Phys. Rev. Appl.},
  volume    = {23},
  issue     = {2},
  pages     = {L021002},
  numpages  = {6},
  year      = {2025},
  month     = feb,
  publisher = {American Physical Society},
  doi       = {10.1103/PhysRevApplied.23.L021002},
  url       = {https://link.aps.org/doi/10.1103/PhysRevApplied.23.L021002}
}

@article{Creedon2016,
  author  = {Creedon, Daniel L. and Goryachev, Maxim and Kostylev, Nikita and Sercombe, Timothy B. and Tobar, Michael E.},
  title   = {A {3D} printed superconducting aluminium microwave cavity},
  journal = {Applied Physics Letters},
  volume  = {109},
  number  = {3},
  pages   = {032601},
  year    = {2016},
  month   = jul,
  issn    = {0003-6951},
  doi     = {10.1063/1.4958684},
  url     = {https://doi.org/10.1063/1.4958684}
}

@article{Krupka2005,
  author  = {J. Krupka and M. E. Tobar and J. G. Hartnett and D. Cros and J.-M. Le Floch},
  title   = {Extremely high-Q factor dielectric resonators for millimeter-wave applications},
  journal = {IEEE Trans. Microw. Theory Techn.},
  volume  = {53},
  number  = {2},
  pages   = {702--712},
  year    = {2005},
  doi     = {10.1109/TMTT.2004.841697}
}

@misc{Gopalakrishnan2024,
  title         = {Guided modes of helical waveguides},
  author        = {Jay Gopalakrishnan and Michael Neunteufel},
  year          = {2025},
  eprint        = {2506.03276},
  archiveprefix = {arXiv},
  primaryclass  = {physics.optics},
  url           = {https://arxiv.org/abs/2506.03276}
}

@article{Piggott2017FabricationConstrained,
  author    = {Piggott, Alexander Y. and Petykiewicz, Jan A. and Su, Logan and Vučković, Jelena},
  title     = {Fabrication-constrained nanophotonic inverse design},
  journal   = {Scientific Reports},
  year      = {2017},
  volume    = {7},
  pages     = {1786},
  doi       = {10.1038/s41598-017-01939-2},
  url       = {https://doi.org/10.1038/s41598-017-01939-2},
  issn      = {2045-2322},
  publisher = {Nature Publishing Group}
}

@article{InverseDesigned_2021_Vercruysse,
  author  = {Vercruysse, Dries and Sapra, Neil V. and Yang, Ki Youl and Vučković, Jelena},
  title   = {Inverse-Designed Photonic Crystal Circuits for Optical Beam Steering},
  journal = {ACS Photonics},
  volume  = {8},
  number  = {10},
  pages   = {3085--3093},
  year    = {2021},
  doi     = {10.1021/acsphotonics.1c01119},
  url     = {https://doi.org/10.1021/acsphotonics.1c01119},
  eprint  = {https://doi.org/10.1021/acsphotonics.1c01119}
}

@article{Beam_Stearing_ID_Thureja_2020,
  author  = {Thureja, Prachi and Shirmanesh, Ghazaleh Kafaie and Fountaine, Katherine T. and Sokhoyan, Ruzan and Grajower, Meir and Atwater, Harry A.},
  title   = {Array-Level Inverse Design of Beam Steering Active Metasurfaces},
  journal = {ACS Nano},
  volume  = {14},
  number  = {11},
  pages   = {15042--15055},
  year    = {2020},
  doi     = {10.1021/acsnano.0c05026},
  url     = {https://doi.org/10.1021/acsnano.0c05026}
}

@article{ZolfagharyPour2026,
  author  = {Zolfaghary Pour, Saeed and Zhang, Hanxiang and Liu, Po Wei and Arigong, Bayaner},
  title   = {Inverse Design and {3D} Printing of a Multiport Microwave Power Splitter: A Scalable Electromagnetic Design Framework},
  journal = {Communications Engineering},
  year    = {2026},
  volume  = {5},
  pages   = {53},
  doi     = {10.1038/s44172-026-00601-y},
  url     = {https://doi.org/10.1038/s44172-026-00601-y}
}

@article{Morrison2025Rapid,
  title     = {Rapid inverse design of microwave devices with scattering matrices},
  author    = {Morrison, Nathaniel and He, Xujia and Xie, Tianhang and Ma, Eric Y.},
  journal   = {Phys. Rev. Appl.},
  volume    = {24},
  issue     = {4},
  pages     = {044100},
  numpages  = {8},
  year      = {2025},
  month     = oct,
  publisher = {American Physical Society},
  doi       = {10.1103/qlfy-4qhw},
  url       = {https://link.aps.org/doi/10.1103/qlfy-4qhw}
}

@article{Khoram2020,
  author    = {Erfan Khoram and Xiaoping Qian and Ming Yuan and Zongfu Yu},
  journal   = {Opt. Express},
  number    = {5},
  pages     = {7060--7069},
  publisher = {Optica Publishing Group},
  title     = {Controlling the minimal feature sizes in adjoint optimization of nanophotonic devices using b-spline surfaces},
  volume    = {28},
  month     = mar,
  year      = {2020},
  url       = {https://opg.optica.org/oe/abstract.cfm?URI=oe-28-5-7060},
  doi       = {10.1364/OE.384438}
}

@article{Inverse_Design_Demultiplexer_Su_2018,
  author  = {Su, Logan and Piggott, Alexander Y. and Sapra, Neil V. and Petykiewicz, Jan and Vučković, Jelena},
  title   = {Inverse Design and Demonstration of a Compact on-Chip Narrowband Three-Channel Wavelength Demultiplexer},
  journal = {ACS Photonics},
  volume  = {5},
  number  = {2},
  pages   = {301--305},
  year    = {2018},
  doi     = {10.1021/acsphotonics.7b00987},
  url     = {https://doi.org/10.1021/acsphotonics.7b00987}
}

@software{MPH,
  author    = {John Hennig and
               Andreas Maeder and
               Annika Schultheiss and
               Jacob Feder},
  title     = {{MPh}-py/{MPh}: 1.3.0},
  month     = nov,
  year      = 2025,
  publisher = {Zenodo},
  version   = {v1.3.0},
  doi       = {10.5281/zenodo.17555089},
  url       = {https://doi.org/10.5281/zenodo.17555089},
  swhid     = {swh:1:dir:83a1c6c423e4193b423856b64be9ef36ee87523d
               ;origin=https://doi.org/10.5281/zenodo.4685961;vis
               it=swh:1:snp:84c70cb78e322aaf97896809a065726243b34
               81f;anchor=swh:1:rel:b564f922544513a4a943d791fee31
               4575ccc7362;path=MPh-py-MPh-86ffed4}
}

@misc{comsol_multiphysics_v6.4,
  author       = {{COMSOL Inc.}},
  title        = {{COMSOL Multiphysics®}},
  year         = {2024},
  version      = {6.4},
  organization = {COMSOL AB},
  address      = {Stockholm, Sweden},
  url          = {www.comsol.com}
}

\end{document}